\documentclass[10pt,twocolumn,superscriptaddress,floatfix,nobalancelastpage,showpacs,amsmath,longbibliography]{revtex4-1}

\usepackage{array,longtable}
\newcolumntype{L}{>{\tiny $}p{0.33\columnwidth}<{$}}
\newcolumntype{M}{>{\scriptsize $}p{0.33\columnwidth}<{$}}
\newcolumntype{N}{>{\scriptsize $}p{0.43\columnwidth}<{$}}
\usepackage{amsmath}
\usepackage{amssymb}
\usepackage{amsfonts}
\usepackage{graphicx}
\usepackage{hyperref}
\usepackage{mathbbol}
\usepackage[vcentermath]{youngtab}

\usepackage{float}
\usepackage{graphics}

\allowdisplaybreaks[1]

\newif\ifhyper
\hypertrue
\ifhyper
\hypersetup{
   citecolor = {green},
   colorlinks = {true}, 
   urlcolor = {blue} 
}

\DeclareMathOperator{\Tr}{Tr}

\begin{document}


\title{
$SU(4)$ topological RVB spin liquid on the square lattice
}

%
\author{Olivier Gauth\'e}
\affiliation{Laboratoire de Physique Th\'eorique, Universit\'e de Toulouse, CNRS, UPS, France}
\author{Sylvain Capponi}
\affiliation{Laboratoire de Physique Th\'eorique, Universit\'e de Toulouse, CNRS, UPS, France}
\author{Didier Poilblanc}
\affiliation{Laboratoire de Physique Th\'eorique, Universit\'e de Toulouse, CNRS, UPS, France}
\affiliation{Institute of Theoretical Physics, \'Ecole Polytechnique F\'ed\'erale de Lausanne (EPFL), CH-1015 Lausanne, Switzerland}


\date{\today}


\begin{abstract}
We generalize the construction of the spin-1/2 SU(2) Resonating Valence Bond (RVB) state to the case of the self-conjugate $\textbf{6}$-representation of SU(4).
As for the case of SU(2) [J-Y. Chen and D. Poilblanc, Phys. Rev. B {\bf 97}, 161107(R) (2018)], we use the Projected Entangled Pair State (PEPS) formalism to derive a simple (two-dimensional) family of generalized SU(4) RVB states
on the square lattice. We show that, when longer-range SU(4)-singlet bonds are included, a local gauge symmetry is broken down from U(1) to $\mathbb{Z}_2$, leading to the emergence 
of a short-range spin liquid. Evidence for the topological nature of this spin liquid is provided  by the investigation of the Renyi entanglement entropy of infinitely-long cylinders and of the modular matrices. Relevance to microscopic models and experiments of ultracold atoms is discussed.
\end{abstract}

\maketitle

{\it Introduction --}
Physical systems such as ultracold atoms loaded on two-dimensional (2D) optical lattices~\cite{bloch_ultracold_2005, lewenstein_ultracold_2012, cazalilla_ultracold_2014}  are ruled by SU(N) symmetry and offer a rich variety of phase diagrams both experimentally
and theoretically~\cite{cazalilla_ultracold_2009, laflamme_c_2016, capponi_phases_2016}. They constitute a rich playground to investigate quantum spin liquid (SL) phases, where the SU(N) symmetry is not broken~\cite{wen_sl_2002}. 

\begin{figure}
	\centering
		\includegraphics[width=0.45\textwidth]{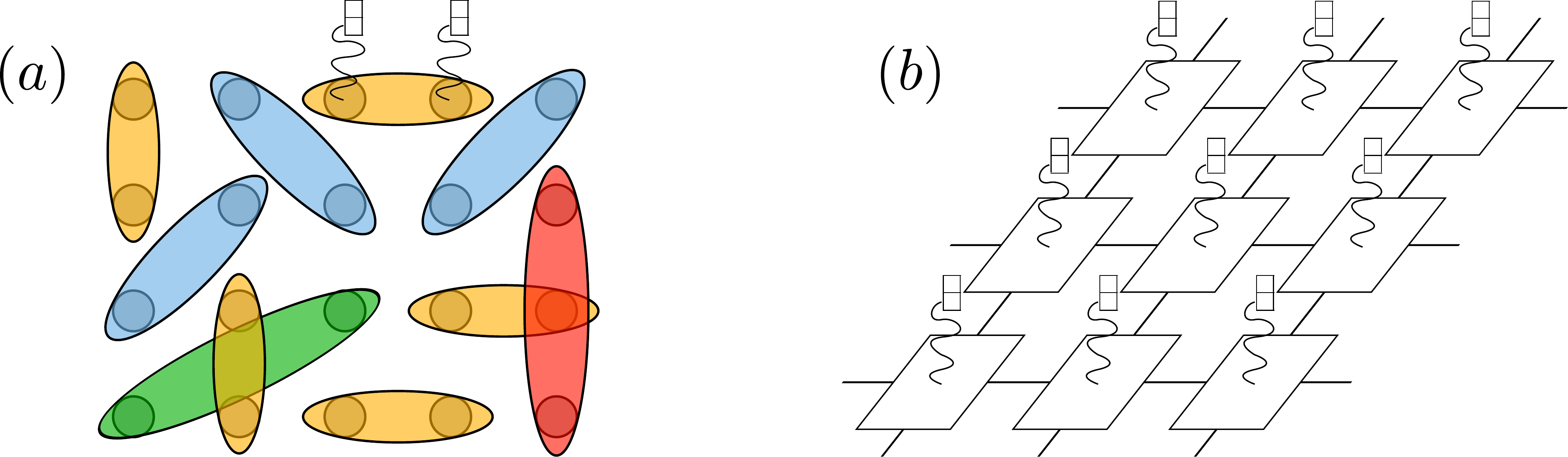}
		\caption{\footnotesize{$(a)$ RVB state defined as a resonance state between SU(4) singlet bond configurations on a 2D square lattice. 
		Singlets (shown as ellipses) are made from two antisymmetric {\bf 6}-representations (represented as two box Young tableaus). $(b)$ SU(4) RVB state represented as a 2D PEPS tensor network.}}
		\label{fig:rvb}
\end{figure}

Tensor network (TN) algorithms such as those using the framework of Projected Entangled Pair States (PEPS) have proven extremely efficient to simulate various quantum SU(2) spin models~\cite{poilblanc_su2_1_2016, poilblanc_quantum_2017, poilblanc_investigation_2017, niesen_tensor_2017}. When dealing with spin liquids, it is possible to encode {\it global} spin-rotation invariance and lattice symmetries directly at the level of a unique site
tensor~\cite{mambrini_systematic_2016, schmoll_programming_2018} which completely characterizes the quantum state. In addition, topological order can also be encoded via a {\it local} gauge symmetry of the tensor~\cite{schuch_topo_2010}. These remarkable
properties make the PEPS formalism particularly suited to construct simple Ans\"atze of topological SLs. For example, the Resonating Valence Bond (RVB) state~\cite{anderson_resonating_1987} can be written as a PEPS~\cite{schuch_resonating_2012} and was proven to be a topological SL belonging to the same $\mathbb{Z}_2$ topological class as the Toric Code~\cite{kitaev_anyons_2006} (TC),
both on the Kagome~\cite{poilblanc_topological_2012} and on the square lattice~\cite{chen_topological_2018}. Although the construction can readily be exported to SU($N$) for $N>2$, it has mostly be applied
so far to derive a simple Affleck-Kennedy-Lieb-Tasaki (AKLT) trivial (i.e. non-topological) SU(3) liquid~\cite{gauthe_entanglement_2017}, as well as a few topological SU(3) liquids~\cite{kurecic_gapped_2019, dong_su3_2018}. We here construct a family of simple $\mathbb{Z}_2$ topological SU(4) liquids on the square lattice. Our construction is similar to the PEPS construction of spin-1/2 extended  RVB states~\cite{chen_topological_2018},
replacing the fundamental irreducible representation (irrep) of SU(2), namely the two-dimensional spin-1/2 irrep, by the self-conjugate {\bf 6}-representation of SU(4). This representation can be seen as the Hilbert space of two fermions with four colors; it is represented by a vertical Young tableau of two boxes $\tiny\yng(1,1)$~\cite{kim_linear_2017}.
As for SU(2), the PEPS wavefunction can be viewed as a linear superposition (or "resonance") of short-range singlet bond 
configurations~\cite{anderson_resonating_1987} as seen in Fig.~\ref{fig:rvb} ($a$).
Note that our SU(4) RVB state may not only contains nearest-neighbor (NN) singlet bonds (so-called NN RVB state) but also longer range singlet bonds, including some connecting the {\it same} sublattice.

\begin{figure}
  \centering
    \includegraphics[width=0.45\textwidth]{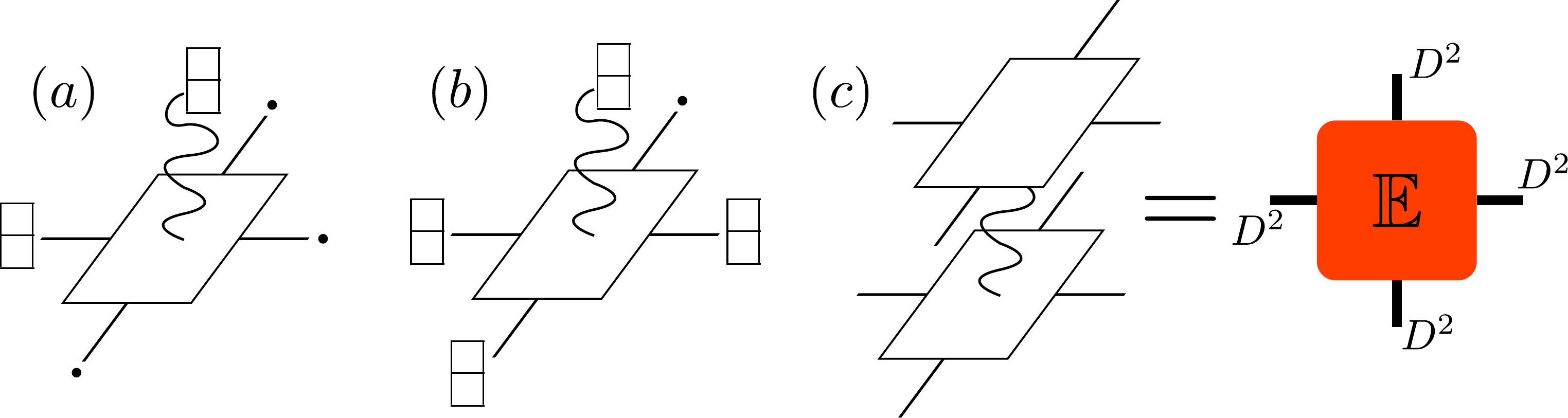}
    \vskip 0.2truecm
    \includegraphics[width=0.45\textwidth]{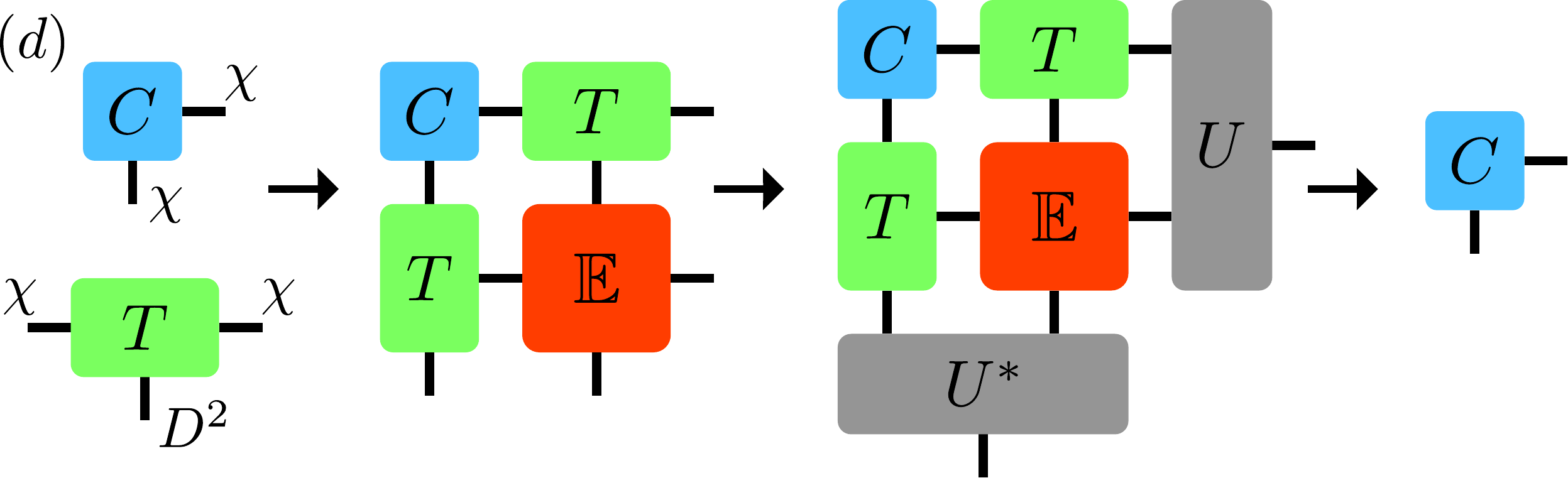}
        \vskip 0.2truecm
    \includegraphics[width=0.35\textwidth]{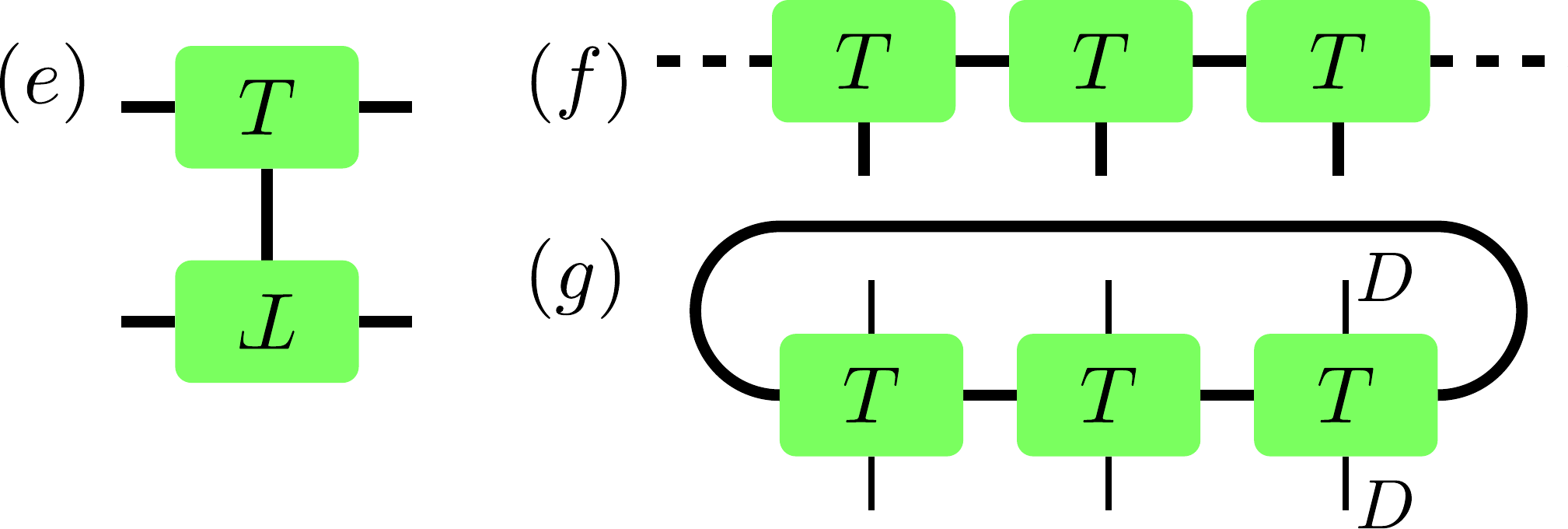}
        \caption{\footnotesize{$(a)$ NN RVB tensor $T_0$ with only one virtual $\bf 6$ state. $(b)$  $T_1, T_2$ and $T_3$ tensors with three virtual $\bf 6$ states. 
$(c)$ In the iPEPS method one constructs first a bi-layer $\mathbb{E}=A\otimes A^*$ tensor by contracting the physical index. 
$(d)$ Corner Transfer Matrix Renormalization Group (CTMRG) algorithm for a rotation invariant tensor $\mathbb{E}$. A unique corner matrix $C$ and a unique side tensor $T$ are renormalized by adding $\mathbb{E}$ iteratively.  
$(e)$ Transfer matrix ${\cal T}=T\otimes T$. 
$(f)$ $|\Psi_{\rm Env}\big>$ (infinite) quantum chain obtained by contracting the $\mathbb{E}$ TN on half of the 2D plane (here from top to bottom).
$(g)$ $\sigma_b$ boundary operator on a $N_v=3$ ring. The $D^2$ degrees of freedom on each site have been reshaped as $D\times D$. 
}}
    \label{fig:tens_SU4}
\end{figure}

{\it PEPS Ansatz --}
In order to generalize the square-lattice RVB state~\cite{chen_topological_2018} to SU(4), one attaches four virtual variables, either $\textbf{6}$ or  $\textbf{1}$ (trivial irrep), to each site along the four 
lattice directions (see Figs.~\ref{fig:tens_SU4}($a$,$b$)), in order to fully cover the square lattice with all possible configurations
of $\bf 6-6$ NN singlet bonds or $\bf 1-1$ NN "empty" bonds. The entangled RVB state is then obtained by projecting the four virtual states into the
physical $\bf 6$-representation, on every site, forming the PEPS $|\Psi_{\rm PEPS}\big>$ represented in \hbox{Fig.~\ref{fig:rvb} (b)}.  In fact, 16 tensors can be constructed  
projecting the virtual space $(\textbf{6} \oplus \textbf{1})^{\otimes 4}$, of dimension $D^4$ with $D=7$, onto the physical SU(4) irrep \textbf{6}, of dimension $d=6$.
Rotation-invariant tensors can be classified according to the $A_1$ and $A_2$ representations of the point group $C_{4v}$. The tensor $T_0$ of Fig.~\ref{fig:tens_SU4}(a) 
representing the NN RVB state is the same as the NN RVB tensor defined for the alternated $\textbf{6}-\overline{\textbf{6}}$ representations of SU(6)\cite{stephan_entanglement_2013} and has a local $A_1$ symmetry.  The tensors $T_1$ and $T_2$ have $A_1$ symmetry and  the tensor $T_3$ has $A_2$ symmetry (see Fig.~\ref{fig:tens_SU4}(b)). The tensors $T_0$ and $T_1$ are actually invariant under any virtual leg permutation ($S_4$ symmetry group). The most general rotation-invariant tensor describing a short-range RVB state can be written 
as
\begin{equation}
A = a_0 T_0 + a_1 T_1 + a_2 T_2 +  i a_3 T_3,
\label{eq:tensor}
\end{equation}
where the imaginary $i$ factor in front of $T_3$ is needed to preserve the full lattice symmetry (at the price of breaking time reversal symmetry). The expressions of these 4 elementary tensors can be found in the supplemental material (SM)~\cite{supp}. 
Since two different singlets can be made on a bond, either \hbox{$\textbf{6} \otimes \textbf{6} \rightarrow \textbf{1}$} or $\textbf{1} \otimes \textbf{1} \rightarrow \textbf{1}$, there is a one-dimensional family of valid bond projector parametrized by an angle $\kappa$,
$
P(\kappa) = \cos\kappa (\mathcal{P}_{\textbf{6}\otimes \textbf{6}} \otimes \mathbb{1}) + \sin\kappa (\mathbb{1} \otimes \mathcal{P}_{\textbf{1}\otimes \textbf{1}})
$,
which are all equivalent up to renormalization of the $T_i$ tensors~\cite{jiang_symmetric_2015} (for simplicity, we chose $\kappa = \pi/4$).
Implicitly, this projector is applied on every bond of the TN (see SM for details~\cite{supp}).

{\it CTMRG algorithm --}
In the following we work directly in the thermodynamic limit, i.e. on an infinite 2D lattice, with the "double-layer" TN representing the norm $\big<\Psi_{\rm PEPS}|\Psi_{\rm PEPS}\big>$ (iPEPS method). The Corner Transfer Matrix Renormalization Group (CTMRG) algorithm~\cite{orus_simulation_2009} is an efficient way to contract the corresponding elementary tensor $\mathbb{E}=A\otimes A^*$ shown in Fig.~\ref{fig:tens_SU4}~(c).
This procedure involves controlled approximations, using a real space renormalization group technique, depicted in Fig.~\ref{fig:tens_SU4}~($d$). The infinite-volume environment of  
any (rectangular) region of space is approximated using  an edge tensor $T$ of dimension
$\chi\times D^2\times\chi$ and a corner $\chi\times\chi$ matrix $C$, where $\chi$ is the environment dimension taken as large as possible. $C$ is obtained iteratively by adding tensors $\mathbb{E}$ to the corner, diagonalizing the resulting hermitian matrix and keeping only $\chi$ eigenvalues. The exact value of $\chi$ has to fit the decomposition of the spectrum into SU(4) multiplets and cannot be fixed arbitrarily. A new tensor $T$ is then computed using the diagonalization basis $U$. Thanks to lattice rotation invariance, the renormalization process can be performed on one corner only; it is processed until the spectrum of $C$
has converged.

By (approximately) contracting the $\mathbb{E}$ tensor network from infinitely far away on the left and right of the 2D plane, one then ends up with an infinitely long (vertical)  two-leg ladder $(T\otimes T)^L$,
$L\rightarrow\infty$ (where the $D^{2L}$ virtual indices between the two chains of $T$ tensors are contracted). Using Lanczos algorithm, one can extract the two leading eigenvalues $\lambda_1$ and $\lambda_2$ 
of the corresponding ladder transfer matrix ${\cal T}$ (see Fig.~\ref{fig:tens_SU4}(e)) and compute the correlation length $\xi$ of the system from the TM gap $\Delta$ as
$
\Delta = \frac{1}{\xi} = \log\left(\frac{\lambda_1}{\lambda_2}\right)
$.
Note that the finiteness of the corner dimension $\chi$ automatically implies a finite correlation length, so that a finite-$\chi$ scaling analysis is necessary. For a short-ranged wavefunction, $\xi(\chi)$ converges exponentially fast towards a finite length $\xi_\infty$.
In contrast, for critical $|\Psi_{\rm PEPS}\big>$ wavefunctions, 
$\xi$ increases linearly with $\chi$, diverging in the relevant $\chi\rightarrow\infty$ limit.

\begin{table}[H]
\caption{Properties of the four critical PEPS. $^*$Tensors invariant under the larger $S_4$ symmetric group (see text).}
\begin{ruledtabular}
\begin{tabular}{cccc}
tensor\; & \;$C_{4v}$ symmetry & \; central charge \; & \; dimer exponent $\alpha$ \\
\hline
   $T_0$ &    $A_1\!^*$    &       1      & $1.75 \pm 0.02$   \\
   $T_1$ &    $A_1\!^*$    &       1      & $1.57 \pm 0.15$   \\
   $T_2$ &    $  A_1$      &       1      & $1.83 \pm 0.08$   \\
   $T_3$ &    $  A_2$      &       1      & $1.45 \pm 0.11$   \\
\end{tabular}
\end{ruledtabular}
\label{table:critical}
\end{table}

{\it Critical spin liquids --}
The four elementary tensors $T_0,T_1, T_2$ and $T_3$ define critical states with long-distance correlations. 
In fact, they can be viewed as Rokhsar-Kilvelson (i.e. equal weight) ground states of quantum dimer models~\cite{rokhsar_superconductivity_1988}, where the "dimers" correspond here either to $\bf 6-6$ (for $T_0$) or to $\bf 1-1$ (for $T_i$, $i=1,2,3$) singlet bonds. On bipartite lattices, such as the square lattice, a simple mapping onto a (coarse-grained) height field theory~\cite{fradkin_rk_2004} implies
algebraically-decaying dimer-dimer correlations $C_d(r)\sim (1/r)^\alpha$ versus distance $r$. Note that the height representation is linked directly to a local
U(1) gauge symmetry of the $A$ tensor. 
We have directly computed the dimer-dimer correlation functions $C_d(r)$ in these states (using the environment $C$ and $T$ tensors above) and extracted the critical exponents $\alpha$ 
by fitting the numerical results~\cite{supp}.  The values of $\alpha$ are listed in Table~\ref{table:critical}. Quite notably, RVB wavefunctions can be related to classical interacting dimer models, thus allowing to estimate their critical correlation exponent~\cite{damle_resonating_2012, alet_classical_2006}. Using 
this equivalence of the $T_0$ PEPS with the SU(6) RVB state, one predicts $\alpha \simeq 1.64$ which is very close to the value we have measured.

\begin{figure}[htbp]
  \centering
    \includegraphics[width=0.4\textwidth]{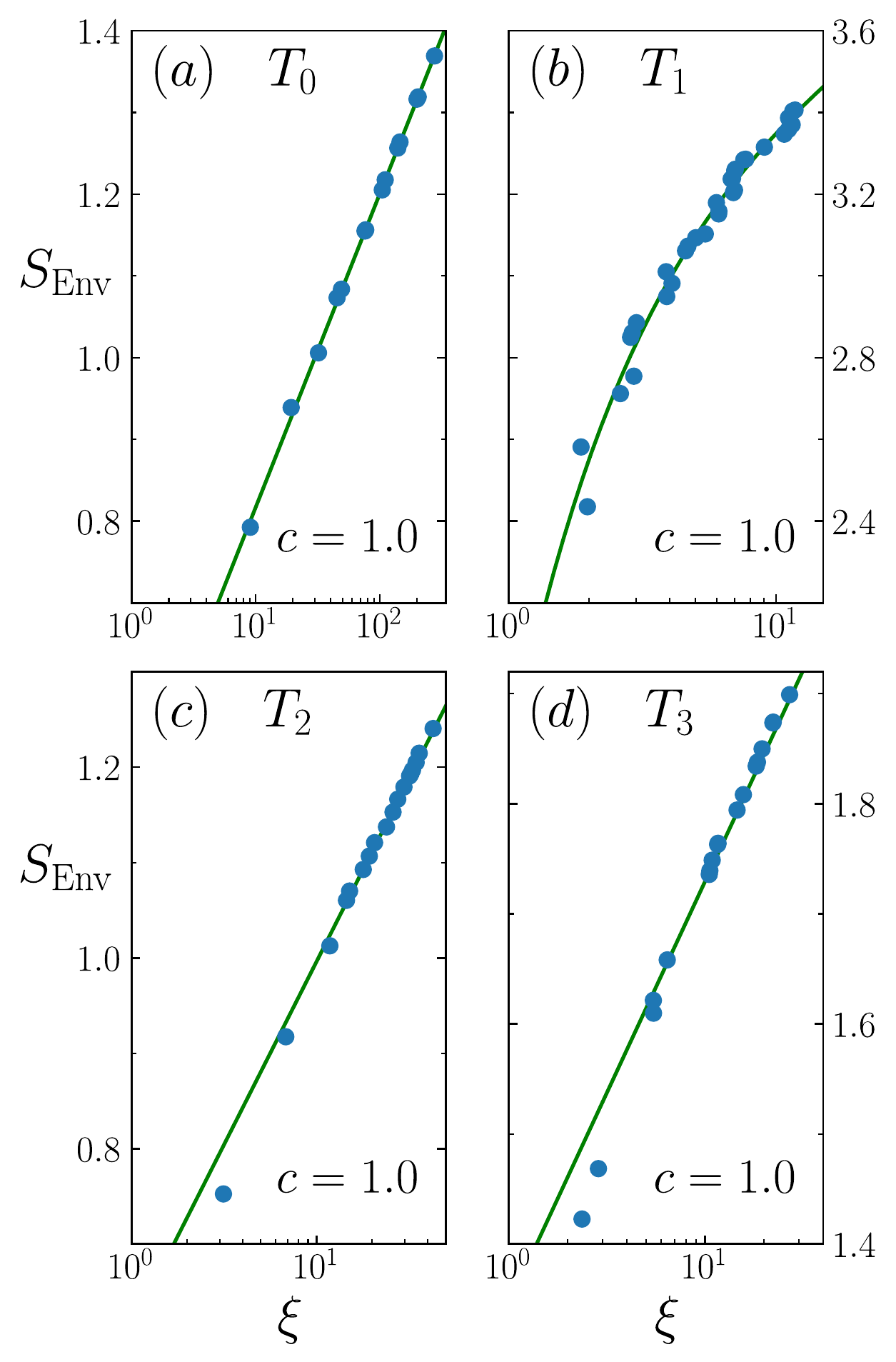}
    \caption{\footnotesize{Environment EE $S_\text{Env}$ plotted as a function of the logarithm of the correlation length $\xi$.
    Linear fits enable to extract the central charge $c$ of the wavefunctions defined by $T_0$, $T_2$ and $T_3$; a higher order correction $a\, \xi^{-b}$ is included for $T_1$ . We find $c \simeq 1$ in all cases.
    }}
    \label{fig:s1T}
\end{figure}

In order to characterize the underlying conformal field theory by its central charge $c$, we 
have investigated the entanglement entropy (EE) of the one-dimensional quantum state $|\Psi_{\rm Env}\big> \in [\mathbb{C}^{D^2}]^{\otimes L}$ made of an infinite ($L\rightarrow\infty$) chain $T^{\otimes L}$ of $\chi$-contracted $T$ tensors (see Fig.~\ref{fig:tens_SU4}~($f$)). The so-called "environment" EE is computed from the leading eigenvector $\Sigma$ 
(reshaped as a $\chi\times \chi$ matrix) of the above $\cal T$ TM, as $S_{\rm Env} = -\Tr \Sigma^2 \log \Sigma^2$. For a critical wavefunction, the relation $S_{\rm Env} = c/6 \log\xi + S_0$~\cite{calabrese_entanglement_2004} allows to compute the central charge $c$ (see Fig.~\ref{fig:s1T}). We found $c=1$ in all cases, consistently with the prediction of the height field theory. For $T_1$, the correlation length stays small even for large $\chi$ and we had to include a higher order correction to the fit.

\begin{figure}
  \centering
    \includegraphics[width=0.42\textwidth]{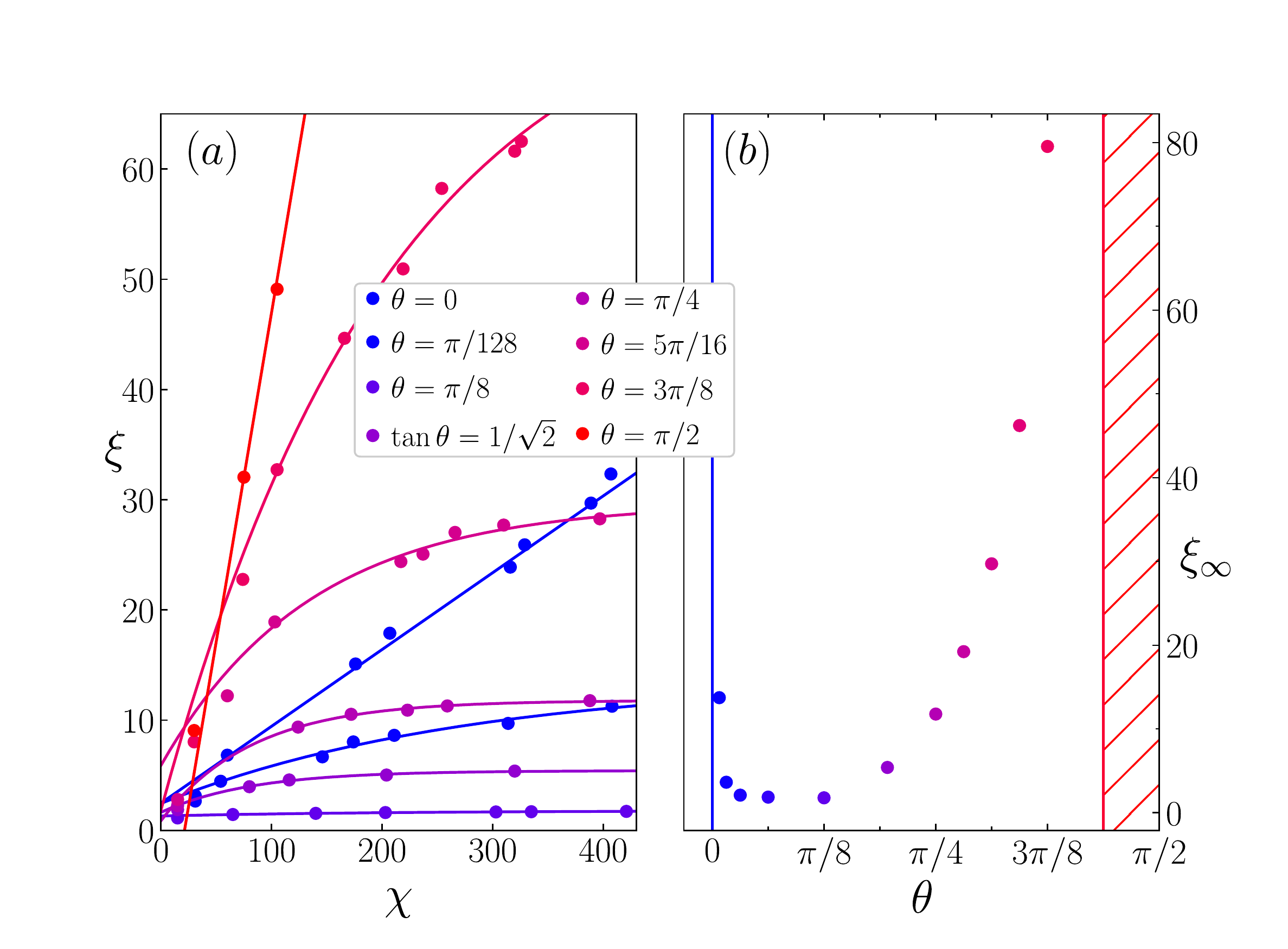}
    \caption{\footnotesize{
    $(a)$ Correlation length versus environment dimension for $\phi=0$, and various values of $\theta$ interpolating $A(\theta,0)$ between $T_2$ ($\theta=0$) and $T_0$ ($\theta=\pi/2$). The $\chi \rightarrow \infty$ correlation length $\xi_\infty$ (for $\theta\ne 0,\pi/2$) is obtained from an exponential fit $\xi(\chi) = \xi_\infty
    + a \exp(-\chi/\ell)$.
    $(b)$  Correlation length $\xi_\infty$ versus $\theta$ (for $\phi=0$).  The correlation length seems to diverge 
    for $\theta<\pi/2$ (in the vicinity of the NN RVB) in the hashed region. In contrast, no critical region was found in the vicinity of the critical point $\theta=\phi=0$ (blue line). Note that the correlation length has a minimum around $\theta=\pi/8$.}}
    \label{fig:xi}
\end{figure}

{\it Phase diagram --}
For simplicity, we shall not consider the $T_1$ tensor any further and restrict to the two-dimensional PEPS family
parametrized by two angles $\theta$ and $\phi$, defining the coefficients of the $A\equiv A(\theta,\phi)$ tensor in Eq.~(\ref{eq:tensor}) as 
$a_0=\cos{\phi}\sin{\theta}$, $a_1=0$, $a_2=\cos{\phi}\cos{\theta}$ and $a_3=\sin{\phi}$.
For each elementary tensor, a fix number of dimers -- either 1 for $T_0$ or 3 for $T_1, T_2$  and $T_3$ -- connect to each site. Hence the number of (fluctuating) dimers cutting a given closed loop is conserved. This means the wavefunction defined by one tensor alone has a U(1) gauge symmetry. For any linear combination of the fundamental tensors with $a_0 \neq 0$, long-range dimers appear and this number of fluctuating dimers is no more constant, but its parity is. Thus U(1) symmetry breaks into $\mathbb{Z}_2$ and we expect the above critical SLs to become unstable. As shown in Fig.~\ref{fig:xi}~($a$),
along the line interpolating between the $T_2$ and $T_0$ tensors, one can extract the correlation length from a fit of $\xi(\chi)$ vs $\chi$. 
As shown in Fig.~\ref{fig:xi}~($b$), we found an extended critical region around the NN RVB state. However, a complete scan of the whole two-dimensional
phase diagram (see SM~\cite{supp} for more data) reveals no sign of other extended critical regions in the vicinity of the critical states 
defined by the $T_2$ and $T_3$ tensors.

\begin{figure}
  \centering
    \includegraphics[width=0.45\textwidth]{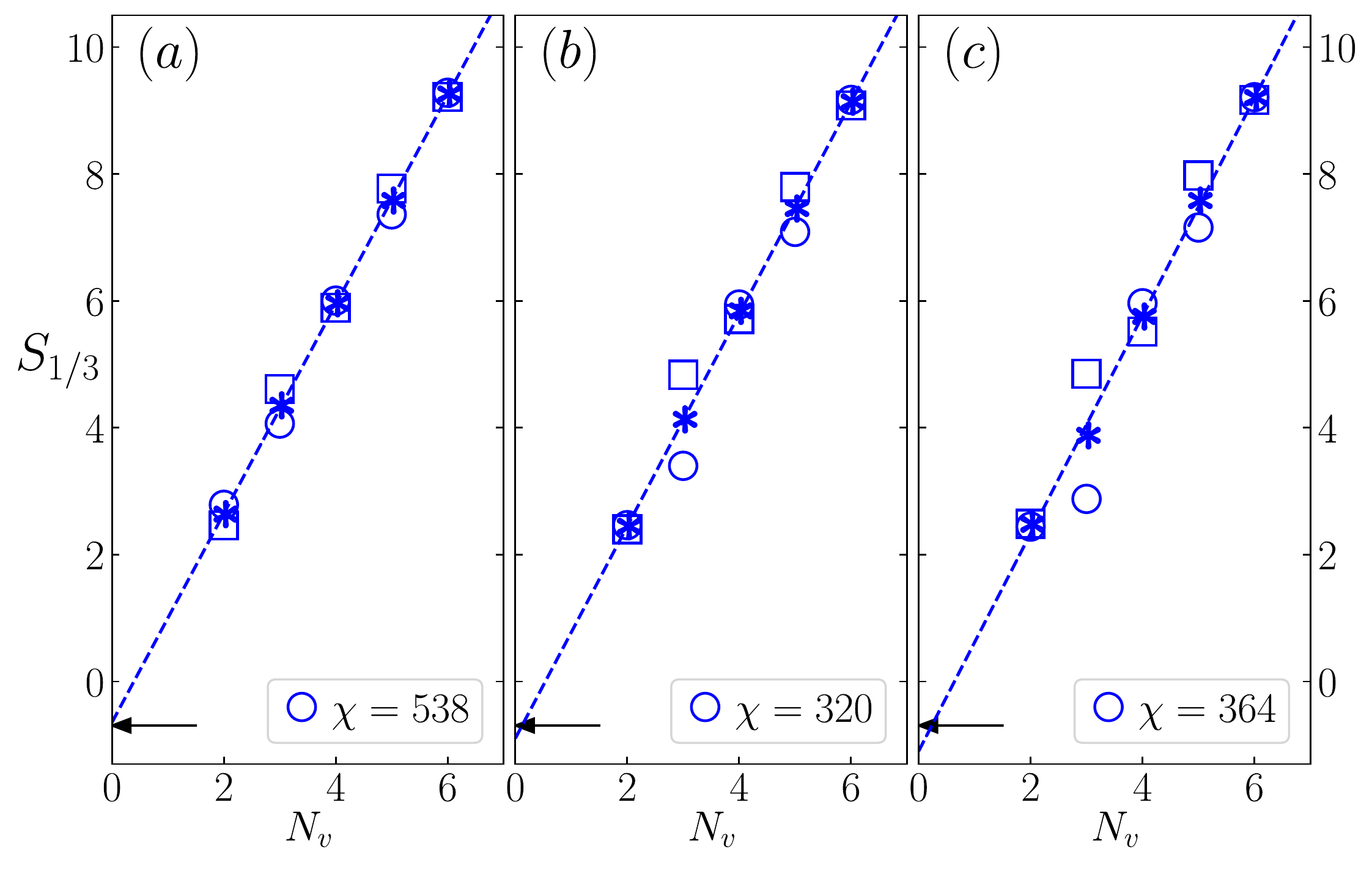}
    \caption{\footnotesize{Renyi entanglement entropies $S_q$ at $q=1/3$ versus the half-cylinder circumference $N_v$ for PEPS parametrized by 
    (a) $\phi=0$ and $\theta=\pi/8$, (b) $\phi=0$ and $\theta=3\pi/8$, and  (c) $\phi=\pi/4$ and $\theta=\pi/2$, obtained at the 
    $\chi$ values indicated on the plots. Squares (circles) label the odd (even) sector. The stars stand for the average of  the two sectors and the dashed lines are linear fits. The arrow points to the value $-\log2$.}}
    \label{fig:s_R3}
\end{figure}

{\it PEPS bulk-edge correspondence and topological EE --}
We now turn to the investigation of the topological properties of the short-range SU(4) SL of our phase diagram. For this, we use the
bulk-edge correspondence theorem of PEPS~\cite{cirac_entanglement_2011} applied to an infinitely long (horizontal) cylinder (of finite perimeter $N_v$) partitioned in two halves 
by a vertical plane~: 
The reduced density matrix $\rho_{\cal C}=\Tr_{\cal C} |\Psi_{\rm PEPS}\big>\big<\Psi_{\rm PEPS}|$ of the half-cylinder ${\cal C}$  can be expressed as $\rho_{\cal C}=U\rho_b U^\dagger$, where $\rho_b$ is a boundary density matrix 
acting on the virtual space $[\mathbb{C}^D]^{\otimes N_v}$ at the boundary
of  ${\cal C}$ and $U$ is an isometry. Naturally, due to the tensor $\mathbb{Z}_2$ gauge symmetry, $\rho_b$ splits into two ${\mathbb Z}_2$-even and ${\mathbb Z}_2$-odd invariant blocks (normalized separately).
Using the CTMRG algorithm, $\rho_b$ can be well approximated by $\rho_b=\sigma_b^2$, where $\sigma_b$ is the product of (reshaped) edge $T$ tensors 
shown in Fig.~\ref{fig:tens_SU4}~($g$), and used to compute Renyi EE defined as $S_q=\frac{1}{1-q} \log \Tr (\rho_b)^q$. Although finite-$\chi$ effects are large for $q>1$,
we have obtained converged results for $q=1/2$ and $q=1/3$~\cite{jiang_accuracy_2013} (see SM~\cite{supp} for details). As shown in Fig.~\ref{fig:s_R3} for three different PEPS, $S_{1/3}$ versus $N_v$ can be well fitted
by a straight line which intersects the vertical axis at a finite value consistent with $-\log2$, the topological EE expected for a $\mathbb{Z}_2$ SL of the same class as 
Kitaev's TC~\cite{kitaev_anyons_2006}. Note the fit in  (b) and (c) may be less reliable since $N_v $ remains smaller than $\xi$. However, in (a) $\xi \ll N_v$.

{\it TRG and modular matrices --}
To obtain further evidence for the topological nature of our family of short-ranged SLs, we have used the Tensor Renormalization Group (TRG) algorithm to compute the
S and T modular matrices~\cite{he_modular_2014}. At the PEPS tensor level, the ${\mathbb Z}_2$ gauge transformation simply amounts to multiplying by $-1$ (respectively $+1$) the {\it virtual} states in the $\textbf{6}$ (respectively $\textbf{1}$) irrep, and this gauge symmetry has to be preserved at each TRG iteration~\cite{he_modular_2014,mei_gapped_2017}. The accuracy of the TRG is also controlled by the maximum number of singular values $\chi$ that are kept at each step, always being careful not to cut multiplets. After convergence, we can easily act with elements of the ${\mathbb Z}_2$ group in order to compute the modular matrices~\cite{chen_topological_2018}.

\begin{figure}
  \centering
    \includegraphics[width=0.35\textwidth]{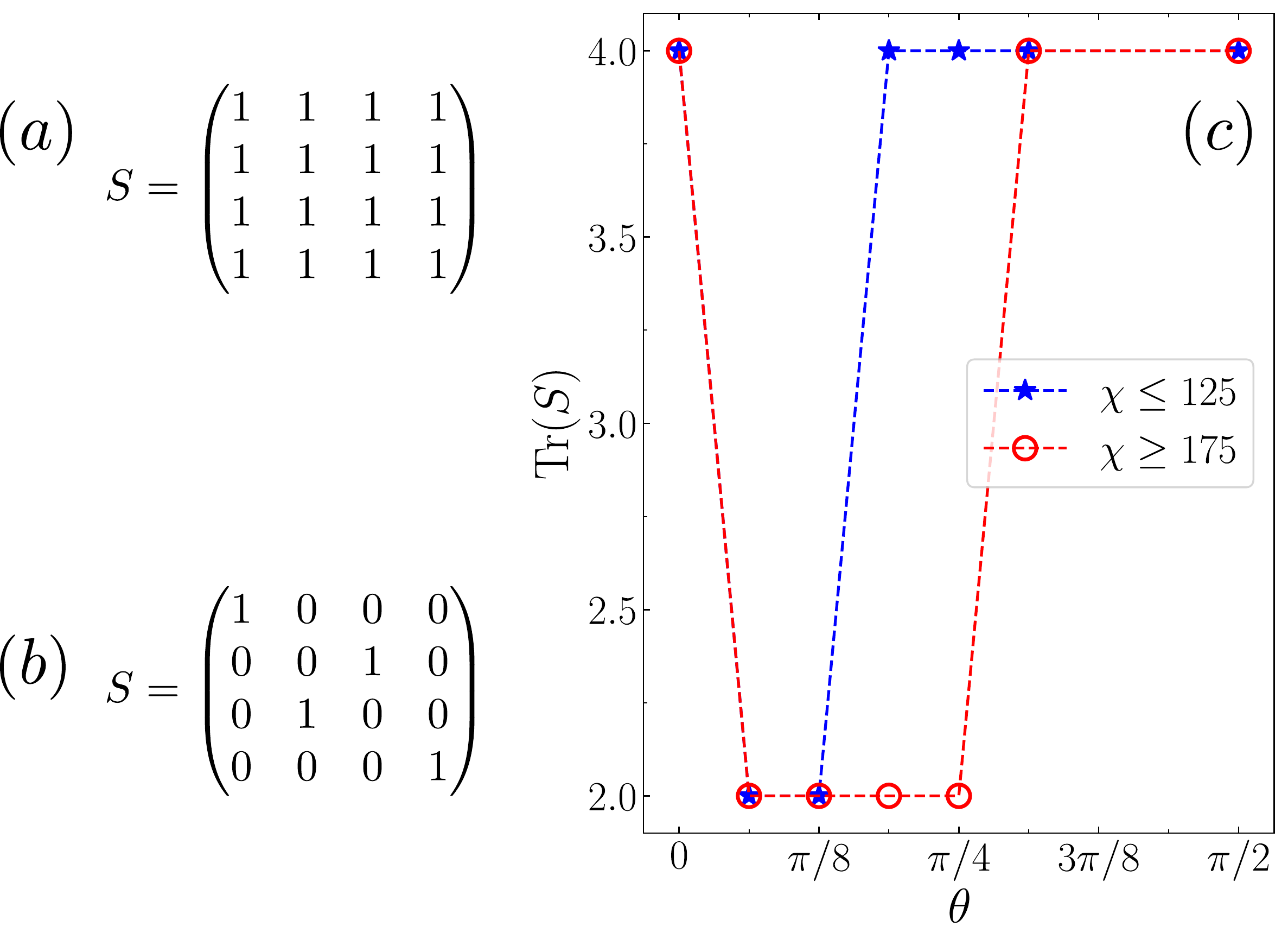}
    \caption{\footnotesize{$(a)$ Modular matrix $S$ of a critical or a  trivial gapped phase. $(b)$ Modular matrix $S$ of a topological phase. $(c)$ Trace of the modular matrix $S$ obtained by TRG for $A(\theta,0)$ versus $\theta$. The value 2 corresponds to a topological phase while 4 implies a trivial phase. We show two sets of data 
   obtained for two ranges of the cut parameter $\chi$ and we note that larger $\chi$ are needed when the correlation length grows.}}
    \label{fig:TRG}
\end{figure}

The two modular matrices $S$ and $T$ are actually linked by permutation of the lines. A critical or a trivial gapped phase (a topological phase with $\mathbb{Z}_2$ topological order) is characterized by the modular matrix $S$ shown in Fig. \ref{fig:TRG}$(a)$ (Fig. \ref{fig:TRG}$(b)$).
The trace of the converged modular matrix $S$ plotted in Fig.~\ref{fig:TRG}$(c)$ vs $\theta$ at fixed $\phi=0$ shows two sharp transitions revealing the existence of  an extended topological SL phase. Note that, when increasing the control parameter $\chi$, the extent of the topological phase gets bigger, in agreement with our previous finding that criticality exists only at $\theta=0$ and in the close vicinity of $\theta=\pi/2$ ($T_0$ PEPS). 

{\it Conclusion --}
In summary, we have extended the construction of the family of spin-1/2 SU(2)-invariant RVB states on the square lattice to a new family of SU(4)-symmetric RVB SLs
with physical $\bf 6$-representation on each site. Physically, such states involve resonance between short-range $\bf 6-6$ singlets, in close analogy with their SU(2) analogs.
Using an exact PEPS representation of these SLs we showed that (i) when singlet dimers are restricted to NN bonds, the SLs are critical, (ii) away from these
fine-tuned states (or small region) the SLs have short-range correlations and (iii) exhibit $\mathbb{Z}_2$ topological order.
By generalizing Oshikawa's argument~\cite{Oshikawa2000}, one can extend Lieb-Schultz-Mattis-Hastings theorem~\cite{Lieb1961,Hastings2004} to conjecture that
it is not possible to obtain a unique featureless gapped state for SU(4) model with physical 6-representation on each site~\footnote{K. Totsuka, private communication. Similarly, a non-degenerate featureless gapped ground state is not possible on the square lattice with SU(4) fundamental and antifundamental representation on the two sublattices~\cite{Jian2018}.}. Note that, for complex PEPS (i.e. when the $T_3$ tensor is included), there is no sign of gapless chiral  edge modes.

We believe our topological SL could be observed both in theoretical models and in experiment. On the theory side, the most general SU(4)-invariant NN Hamiltonian in the $\bf 6$-representation involves only bilinear and biquadratic interactions~\cite{paramekanti2007}.
While a pure bilinear model (Heisenberg like) is expected to stabilize an ordered phase~\cite{kim_linear_2017}, early calculations based on projected wavefunctions~\cite{paramekanti2007}
suggest the existence of a narrow SL region (for a sign of the biquadratic interaction appropriate to a half-filled fermionic Hubbard model~\cite{wang2014}), whose exact nature remains unclear.  Our topological SL is therefore a good candidate for this phase, although other alternatives like a chiral SL - as for the SU(3) triangular lattice~\cite{boos_chiral_2018} - may exist. 
Note that SU(4) differs from SU(2) where no SL survives in the case of NN interactions only. This leaves open the possibility of experimental realizations using any ultracold alkaline-earth atoms realizing SU($N$) symmetry by simply tuning the number of species~\cite{Pagano2014} to e.g. $N=4$. Fixing a filling of two particles per site should avoid three-body losses and thus allow controlled experiments. 

{\it Acknowledgments.}
This   project   is   supported   by   the
TNSTRONG  ANR-16-CE30-0025  and TNTOP ANR-18-CE30-0026-01 grants  (French  Research  Council).  This
work was granted access to the HPC resources of CALMIP
supercomputing   center   under   the   allocation   2018-P1231.
We  acknowledge  inspiring  conversations with Fabien Alet, Ji-Yao Chen, Matthieu Mambrini, Fr\'ed\'eric Mila, Rom\'an Or\'us, David P\'erez-Garc\'ia, Pierre Pujol, and Keisuke Totsuka.
%




\bibliography{bib_SU4_Z2}

\begin{thebibliography}{45}%
\makeatletter
\providecommand \@ifxundefined [1]{%
 \@ifx{#1\undefined}
}%
\providecommand \@ifnum [1]{%
 \ifnum #1\expandafter \@firstoftwo
 \else \expandafter \@secondoftwo
 \fi
}%
\providecommand \@ifx [1]{%
 \ifx #1\expandafter \@firstoftwo
 \else \expandafter \@secondoftwo
 \fi
}%
\providecommand \natexlab [1]{#1}%
\providecommand \enquote  [1]{``#1''}%
\providecommand \bibnamefont  [1]{#1}%
\providecommand \bibfnamefont [1]{#1}%
\providecommand \citenamefont [1]{#1}%
\providecommand \href@noop [0]{\@secondoftwo}%
\providecommand \href [0]{\begingroup \@sanitize@url \@href}%
\providecommand \@href[1]{\@@startlink{#1}\@@href}%
\providecommand \@@href[1]{\endgroup#1\@@endlink}%
\providecommand \@sanitize@url [0]{\catcode `\\12\catcode `\$12\catcode
  `\&12\catcode `\#12\catcode `\^12\catcode `\_12\catcode `\%12\relax}%
\providecommand \@@startlink[1]{}%
\providecommand \@@endlink[0]{}%
\providecommand \url  [0]{\begingroup\@sanitize@url \@url }%
\providecommand \@url [1]{\endgroup\@href {#1}{\urlprefix }}%
\providecommand \urlprefix  [0]{URL }%
\providecommand \Eprint [0]{\href }%
\providecommand \doibase [0]{http://dx.doi.org/}%
\providecommand \selectlanguage [0]{\@gobble}%
\providecommand \bibinfo  [0]{\@secondoftwo}%
\providecommand \bibfield  [0]{\@secondoftwo}%
\providecommand \translation [1]{[#1]}%
\providecommand \BibitemOpen [0]{}%
\providecommand \bibitemStop [0]{}%
\providecommand \bibitemNoStop [0]{.\EOS\space}%
\providecommand \EOS [0]{\spacefactor3000\relax}%
\providecommand \BibitemShut  [1]{\csname bibitem#1\endcsname}%
\let\auto@bib@innerbib\@empty
\bibitem [{\citenamefont {Bloch}(2005)}]{bloch_ultracold_2005}%
  \BibitemOpen
  \bibfield  {author} {\bibinfo {author} {\bibfnamefont {I.}~\bibnamefont
  {Bloch}},\ }\bibfield  {title} {\enquote {\bibinfo {title} {Ultracold quantum
  gases in optical lattices},}\ }\href {\doibase 10.1038/nphys138} {\bibfield
  {journal} {\bibinfo  {journal} {Nature Physics}\ }\textbf {\bibinfo {volume}
  {1}},\ \bibinfo {pages} {23} (\bibinfo {year} {2005})}\BibitemShut {NoStop}%
\bibitem [{\citenamefont {Lewenstein}\ \emph {et~al.}(2012)\citenamefont
  {Lewenstein}, \citenamefont {Sanpera},\ and\ \citenamefont
  {Ahufinger}}]{lewenstein_ultracold_2012}%
  \BibitemOpen
  \bibfield  {author} {\bibinfo {author} {\bibfnamefont {M.}~\bibnamefont
  {Lewenstein}}, \bibinfo {author} {\bibfnamefont {A.}~\bibnamefont {Sanpera}},
  \ and\ \bibinfo {author} {\bibfnamefont {V.}~\bibnamefont {Ahufinger}},\
  }\href
  {http://www.oxfordscholarship.com/view/10.1093/acprof:oso/9780199573127.001.0001/acprof-9780199573127}
  {\emph {\bibinfo {title} {Ultracold {Atoms} in {Optical} {Lattices}:
  {Simulating} quantum many-body systems}}}\ (\bibinfo  {publisher} {Oxford
  University Press},\ \bibinfo {year} {2012})\BibitemShut {NoStop}%
\bibitem [{\citenamefont {Cazalilla}\ and\ \citenamefont
  {Rey}(2014)}]{cazalilla_ultracold_2014}%
  \BibitemOpen
  \bibfield  {author} {\bibinfo {author} {\bibfnamefont {M.~A.}\ \bibnamefont
  {Cazalilla}}\ and\ \bibinfo {author} {\bibfnamefont {A.~M.}\ \bibnamefont
  {Rey}},\ }\bibfield  {title} {\enquote {\bibinfo {title} {Ultracold {Fermi}
  gases with emergent {SU}({N}) symmetry},}\ }\href {\doibase
  10.1088/0034-4885/77/12/124401} {\bibfield  {journal} {\bibinfo  {journal}
  {Reports on Progress in Physics}\ }\textbf {\bibinfo {volume} {77}},\
  \bibinfo {pages} {124401} (\bibinfo {year} {2014})}\BibitemShut {NoStop}%
\bibitem [{\citenamefont {Cazalilla}\ \emph {et~al.}(2009)\citenamefont
  {Cazalilla}, \citenamefont {Ho},\ and\ \citenamefont
  {Ueda}}]{cazalilla_ultracold_2009}%
  \BibitemOpen
  \bibfield  {author} {\bibinfo {author} {\bibfnamefont {M.~A.}\ \bibnamefont
  {Cazalilla}}, \bibinfo {author} {\bibfnamefont {A.~F.}\ \bibnamefont {Ho}}, \
  and\ \bibinfo {author} {\bibfnamefont {M.}~\bibnamefont {Ueda}},\ }\bibfield
  {title} {\enquote {\bibinfo {title} {Ultracold gases of ytterbium:
  ferromagnetism and {Mott} states in an {SU}(6) {Fermi} system},}\ }\href
  {\doibase 10.1088/1367-2630/11/10/103033} {\bibfield  {journal} {\bibinfo
  {journal} {New Journal of Physics}\ }\textbf {\bibinfo {volume} {11}},\
  \bibinfo {pages} {103033} (\bibinfo {year} {2009})}\BibitemShut {NoStop}%
\bibitem [{\citenamefont {Laflamme}\ \emph {et~al.}(2016)\citenamefont
  {Laflamme}, \citenamefont {Evans}, \citenamefont {Dalmonte}, \citenamefont
  {Gerber}, \citenamefont {Mejía-Díaz}, \citenamefont {Bietenholz},
  \citenamefont {Wiese},\ and\ \citenamefont {Zoller}}]{laflamme_c_2016}%
  \BibitemOpen
  \bibfield  {author} {\bibinfo {author} {\bibfnamefont {C.}~\bibnamefont
  {Laflamme}}, \bibinfo {author} {\bibfnamefont {W.}~\bibnamefont {Evans}},
  \bibinfo {author} {\bibfnamefont {M.}~\bibnamefont {Dalmonte}}, \bibinfo
  {author} {\bibfnamefont {U.}~\bibnamefont {Gerber}}, \bibinfo {author}
  {\bibfnamefont {H.}~\bibnamefont {Mejía-Díaz}}, \bibinfo {author}
  {\bibfnamefont {W.}~\bibnamefont {Bietenholz}}, \bibinfo {author}
  {\bibfnamefont {U.-J.}\ \bibnamefont {Wiese}}, \ and\ \bibinfo {author}
  {\bibfnamefont {P.}~\bibnamefont {Zoller}},\ }\bibfield  {title} {\enquote
  {\bibinfo {title} {${CP}({N}-1)$ quantum field theories with alkaline-earth
  atoms in optical lattices},}\ }\href {\doibase 10.1016/j.aop.2016.03.012}
  {\bibfield  {journal} {\bibinfo  {journal} {Annals of Physics}\ }\textbf
  {\bibinfo {volume} {370}},\ \bibinfo {pages} {117} (\bibinfo {year}
  {2016})}\BibitemShut {NoStop}%
\bibitem [{\citenamefont {Capponi}\ \emph {et~al.}(2016)\citenamefont
  {Capponi}, \citenamefont {Lecheminant},\ and\ \citenamefont
  {Totsuka}}]{capponi_phases_2016}%
  \BibitemOpen
  \bibfield  {author} {\bibinfo {author} {\bibfnamefont {S.}~\bibnamefont
  {Capponi}}, \bibinfo {author} {\bibfnamefont {P.}~\bibnamefont
  {Lecheminant}}, \ and\ \bibinfo {author} {\bibfnamefont {K.}~\bibnamefont
  {Totsuka}},\ }\bibfield  {title} {\enquote {\bibinfo {title} {Phases of
  one-dimensional {SU}({N}) cold atomic {Fermi} gases—{From} molecular
  {Luttinger} liquids to topological phases},}\ }\href {\doibase
  10.1016/j.aop.2016.01.011} {\bibfield  {journal} {\bibinfo  {journal} {Annals
  of Physics}\ }\textbf {\bibinfo {volume} {367}},\ \bibinfo {pages} {50}
  (\bibinfo {year} {2016})}\BibitemShut {NoStop}%
\bibitem [{\citenamefont {Wen}(2002)}]{wen_sl_2002}%
  \BibitemOpen
  \bibfield  {author} {\bibinfo {author} {\bibfnamefont {X.-G.}\ \bibnamefont
  {Wen}},\ }\bibfield  {title} {\enquote {\bibinfo {title} {Quantum orders and
  symmetric spin liquids},}\ }\href {\doibase 10.1103/PhysRevB.65.165113}
  {\bibfield  {journal} {\bibinfo  {journal} {Phys. Rev. B}\ }\textbf {\bibinfo
  {volume} {65}},\ \bibinfo {pages} {165113} (\bibinfo {year}
  {2002})}\BibitemShut {NoStop}%
\bibitem [{\citenamefont {Poilblanc}\ \emph {et~al.}(2016)\citenamefont
  {Poilblanc}, \citenamefont {Schuch},\ and\ \citenamefont
  {Affleck}}]{poilblanc_su2_1_2016}%
  \BibitemOpen
  \bibfield  {author} {\bibinfo {author} {\bibfnamefont {D.}~\bibnamefont
  {Poilblanc}}, \bibinfo {author} {\bibfnamefont {N.}~\bibnamefont {Schuch}}, \
  and\ \bibinfo {author} {\bibfnamefont {Ian}\ \bibnamefont {Affleck}},\
  }\bibfield  {title} {\enquote {\bibinfo {title} {${SU(2)}_1$ chiral edge
  modes of a critical spin liquid},}\ }\href {\doibase
  10.1103/PhysRevB.93.174414} {\bibfield  {journal} {\bibinfo  {journal} {Phys.
  Rev. B}\ }\textbf {\bibinfo {volume} {93}},\ \bibinfo {pages} {174414}
  (\bibinfo {year} {2016})}\BibitemShut {NoStop}%
\bibitem [{\citenamefont {Poilblanc}\ and\ \citenamefont
  {Mambrini}(2017)}]{poilblanc_quantum_2017}%
  \BibitemOpen
  \bibfield  {author} {\bibinfo {author} {\bibfnamefont {D.}~\bibnamefont
  {Poilblanc}}\ and\ \bibinfo {author} {\bibfnamefont {M.}~\bibnamefont
  {Mambrini}},\ }\bibfield  {title} {\enquote {\bibinfo {title} {Quantum
  critical phase with infinite projected entangled paired states},}\ }\href
  {\doibase 10.1103/PhysRevB.96.014414} {\bibfield  {journal} {\bibinfo
  {journal} {Phys. Rev. B}\ }\textbf {\bibinfo {volume} {96}},\ \bibinfo
  {pages} {014414} (\bibinfo {year} {2017})}\BibitemShut {NoStop}%
\bibitem [{\citenamefont {Poilblanc}(2017)}]{poilblanc_investigation_2017}%
  \BibitemOpen
  \bibfield  {author} {\bibinfo {author} {\bibfnamefont {D.}~\bibnamefont
  {Poilblanc}},\ }\bibfield  {title} {\enquote {\bibinfo {title} {Investigation
  of the chiral antiferromagnetic {Heisenberg} model using projected entangled
  pair states},}\ }\href {\doibase 10.1103/PhysRevB.96.121118} {\bibfield
  {journal} {\bibinfo  {journal} {Phys. Rev. B}\ }\textbf {\bibinfo {volume}
  {96}},\ \bibinfo {pages} {121118} (\bibinfo {year} {2017})}\BibitemShut
  {NoStop}%
\bibitem [{\citenamefont {Niesen}\ and\ \citenamefont
  {Corboz}(2017)}]{niesen_tensor_2017}%
  \BibitemOpen
  \bibfield  {author} {\bibinfo {author} {\bibfnamefont {I.}~\bibnamefont
  {Niesen}}\ and\ \bibinfo {author} {\bibfnamefont {P.}~\bibnamefont
  {Corboz}},\ }\bibfield  {title} {\enquote {\bibinfo {title} {A tensor network
  study of the complete ground state phase diagram of the spin-1
  bilinear-biquadratic {Heisenberg} model on the square lattice},}\ }\href
  {\doibase 10.21468/SciPostPhys.3.4.030} {\bibfield  {journal} {\bibinfo
  {journal} {SciPost Physics}\ }\textbf {\bibinfo {volume} {3}},\ \bibinfo
  {pages} {030} (\bibinfo {year} {2017})}\BibitemShut {NoStop}%
\bibitem [{\citenamefont {Mambrini}\ \emph {et~al.}(2016)\citenamefont
  {Mambrini}, \citenamefont {Or\'us},\ and\ \citenamefont
  {Poilblanc}}]{mambrini_systematic_2016}%
  \BibitemOpen
  \bibfield  {author} {\bibinfo {author} {\bibfnamefont {M.}~\bibnamefont
  {Mambrini}}, \bibinfo {author} {\bibfnamefont {R.}~\bibnamefont {Or\'us}}, \
  and\ \bibinfo {author} {\bibfnamefont {D.}~\bibnamefont {Poilblanc}},\
  }\bibfield  {title} {\enquote {\bibinfo {title} {Systematic construction of
  spin liquids on the square lattice from tensor networks with {SU}(2)
  symmetry},}\ }\href {\doibase 10.1103/PhysRevB.94.205124} {\bibfield
  {journal} {\bibinfo  {journal} {Phys. Rev. B}\ }\textbf {\bibinfo {volume}
  {94}},\ \bibinfo {pages} {205124} (\bibinfo {year} {2016})}\BibitemShut
  {NoStop}%
\bibitem [{\citenamefont {Schmoll}\ \emph {et~al.}(2018)\citenamefont
  {Schmoll}, \citenamefont {Singh}, \citenamefont {Rizzi},\ and\ \citenamefont
  {Or\'us}}]{schmoll_programming_2018}%
  \BibitemOpen
  \bibfield  {author} {\bibinfo {author} {\bibfnamefont {P.}~\bibnamefont
  {Schmoll}}, \bibinfo {author} {\bibfnamefont {S.}~\bibnamefont {Singh}},
  \bibinfo {author} {\bibfnamefont {M.}~\bibnamefont {Rizzi}}, \ and\ \bibinfo
  {author} {\bibfnamefont {R.}~\bibnamefont {Or\'us}},\ }\bibfield  {title}
  {\enquote {\bibinfo {title} {A programming guide for tensor networks with
  global ${SU}(2)$ symmetry},}\ }\href {http://arxiv.org/abs/1809.08180}
  {\bibfield  {journal} {\bibinfo  {journal} {arXiv:1809.08180}\ } (\bibinfo
  {year} {2018})}\BibitemShut {NoStop}%
\bibitem [{\citenamefont {Schuch}\ \emph {et~al.}(2010)\citenamefont {Schuch},
  \citenamefont {Cirac},\ and\ \citenamefont
  {P\'erez-Garc\'ia}}]{schuch_topo_2010}%
  \BibitemOpen
  \bibfield  {author} {\bibinfo {author} {\bibfnamefont {N.}~\bibnamefont
  {Schuch}}, \bibinfo {author} {\bibfnamefont {I.}~\bibnamefont {Cirac}}, \
  and\ \bibinfo {author} {\bibfnamefont {D.}~\bibnamefont {P\'erez-Garc\'ia}},\
  }\bibfield  {title} {\enquote {\bibinfo {title} {{PEPS} as ground states:
  Degeneracy and topology},}\ }\href {\doibase
  https://doi.org/10.1016/j.aop.2010.05.008} {\bibfield  {journal} {\bibinfo
  {journal} {Annals of Physics}\ }\textbf {\bibinfo {volume} {325}},\ \bibinfo
  {pages} {2153} (\bibinfo {year} {2010})}\BibitemShut {NoStop}%
\bibitem [{\citenamefont {Anderson}(1987)}]{anderson_resonating_1987}%
  \BibitemOpen
  \bibfield  {author} {\bibinfo {author} {\bibfnamefont {P.~W.}\ \bibnamefont
  {Anderson}},\ }\bibfield  {title} {\enquote {\bibinfo {title} {The
  {Resonating} {Valence} {Bond} state in {La}$_2${Cu}{O}$_4$ and
  superconductivity},}\ }\href {\doibase 10.1126/science.235.4793.1196}
  {\bibfield  {journal} {\bibinfo  {journal} {Science}\ }\textbf {\bibinfo
  {volume} {235}},\ \bibinfo {pages} {1196} (\bibinfo {year}
  {1987})}\BibitemShut {NoStop}%
\bibitem [{\citenamefont {Schuch}\ \emph {et~al.}(2012)\citenamefont {Schuch},
  \citenamefont {Poilblanc}, \citenamefont {Cirac},\ and\ \citenamefont
  {P\'erez-Garc\'ia}}]{schuch_resonating_2012}%
  \BibitemOpen
  \bibfield  {author} {\bibinfo {author} {\bibfnamefont {N.}~\bibnamefont
  {Schuch}}, \bibinfo {author} {\bibfnamefont {D.}~\bibnamefont {Poilblanc}},
  \bibinfo {author} {\bibfnamefont {J.~I.}\ \bibnamefont {Cirac}}, \ and\
  \bibinfo {author} {\bibfnamefont {D.}~\bibnamefont {P\'erez-Garc\'ia}},\
  }\bibfield  {title} {\enquote {\bibinfo {title} {Resonating valence bond
  states in the {PEPS} formalism},}\ }\href {\doibase
  10.1103/PhysRevB.86.115108} {\bibfield  {journal} {\bibinfo  {journal} {Phys.
  Rev. B}\ }\textbf {\bibinfo {volume} {86}},\ \bibinfo {pages} {115108}
  (\bibinfo {year} {2012})}\BibitemShut {NoStop}%
\bibitem [{\citenamefont {Kitaev}(2006)}]{kitaev_anyons_2006}%
  \BibitemOpen
  \bibfield  {author} {\bibinfo {author} {\bibfnamefont {A.}~\bibnamefont
  {Kitaev}},\ }\bibfield  {title} {\enquote {\bibinfo {title} {Anyons in an
  exactly solved model and beyond},}\ }\href {\doibase
  10.1016/j.aop.2005.10.005} {\bibfield  {journal} {\bibinfo  {journal} {Annals
  of Physics}\ }\textbf {\bibinfo {volume} {321}},\ \bibinfo {pages} {2}
  (\bibinfo {year} {2006})}\BibitemShut {NoStop}%
\bibitem [{\citenamefont {Poilblanc}\ \emph {et~al.}(2012)\citenamefont
  {Poilblanc}, \citenamefont {Schuch}, \citenamefont {P\'erez-Garc\'ia},\ and\
  \citenamefont {Cirac}}]{poilblanc_topological_2012}%
  \BibitemOpen
  \bibfield  {author} {\bibinfo {author} {\bibfnamefont {D.}~\bibnamefont
  {Poilblanc}}, \bibinfo {author} {\bibfnamefont {N.}~\bibnamefont {Schuch}},
  \bibinfo {author} {\bibfnamefont {David}\ \bibnamefont {P\'erez-Garc\'ia}}, \
  and\ \bibinfo {author} {\bibfnamefont {J.~I.}\ \bibnamefont {Cirac}},\
  }\bibfield  {title} {\enquote {\bibinfo {title} {Topological and entanglement
  properties of resonating valence bond wave functions},}\ }\href {\doibase
  10.1103/PhysRevB.86.014404} {\bibfield  {journal} {\bibinfo  {journal} {Phys.
  Rev. B}\ }\textbf {\bibinfo {volume} {86}},\ \bibinfo {pages} {014404}
  (\bibinfo {year} {2012})}\BibitemShut {NoStop}%
\bibitem [{\citenamefont {Chen}\ and\ \citenamefont
  {Poilblanc}(2018)}]{chen_topological_2018}%
  \BibitemOpen
  \bibfield  {author} {\bibinfo {author} {\bibfnamefont {J.-Y.}\ \bibnamefont
  {Chen}}\ and\ \bibinfo {author} {\bibfnamefont {D.}~\bibnamefont
  {Poilblanc}},\ }\bibfield  {title} {\enquote {\bibinfo {title} {Topological
  $\mathbb{Z}_2$ resonating-valence-bond spin liquid on the square lattice},}\
  }\href {\doibase 10.1103/PhysRevB.97.161107} {\bibfield  {journal} {\bibinfo
  {journal} {Phys. Rev. B}\ }\textbf {\bibinfo {volume} {97}},\ \bibinfo
  {pages} {161107} (\bibinfo {year} {2018})}\BibitemShut {NoStop}%
\bibitem [{\citenamefont {Gauth\'e}\ and\ \citenamefont
  {Poilblanc}(2017)}]{gauthe_entanglement_2017}%
  \BibitemOpen
  \bibfield  {author} {\bibinfo {author} {\bibfnamefont {O.}~\bibnamefont
  {Gauth\'e}}\ and\ \bibinfo {author} {\bibfnamefont {D.}~\bibnamefont
  {Poilblanc}},\ }\bibfield  {title} {\enquote {\bibinfo {title} {Entanglement
  properties of the two-dimensional {SU}(3) {Affleck}-{Kennedy}-{Lieb}-{Tasaki}
  state},}\ }\href {\doibase 10.1103/PhysRevB.96.121115} {\bibfield  {journal}
  {\bibinfo  {journal} {Phys. Rev. B}\ }\textbf {\bibinfo {volume} {96}},\
  \bibinfo {pages} {121115} (\bibinfo {year} {2017})}\BibitemShut {NoStop}%
\bibitem [{\citenamefont {Kure\v{c}i\'c}\ \emph {et~al.}(2019)\citenamefont
  {Kure\v{c}i\'c}, \citenamefont {Vanderstraeten},\ and\ \citenamefont
  {Schuch}}]{kurecic_gapped_2019}%
  \BibitemOpen
  \bibfield  {author} {\bibinfo {author} {\bibfnamefont {I.}~\bibnamefont
  {Kure\v{c}i\'c}}, \bibinfo {author} {\bibfnamefont {L.}~\bibnamefont
  {Vanderstraeten}}, \ and\ \bibinfo {author} {\bibfnamefont {N.}~\bibnamefont
  {Schuch}},\ }\bibfield  {title} {\enquote {\bibinfo {title} {Gapped {SU}(3)
  spin liquid with $\mathbb{Z}_3$ topological order},}\ }\href {\doibase
  10.1103/PhysRevB.99.045116} {\bibfield  {journal} {\bibinfo  {journal} {Phys.
  Rev. B}\ }\textbf {\bibinfo {volume} {99}},\ \bibinfo {pages} {045116}
  (\bibinfo {year} {2019})}\BibitemShut {NoStop}%
\bibitem [{\citenamefont {Dong}\ \emph {et~al.}(2018)\citenamefont {Dong},
  \citenamefont {Chen},\ and\ \citenamefont {Tu}}]{dong_su3_2018}%
  \BibitemOpen
  \bibfield  {author} {\bibinfo {author} {\bibfnamefont {X.-Y.}\ \bibnamefont
  {Dong}}, \bibinfo {author} {\bibfnamefont {J.-Y.}\ \bibnamefont {Chen}}, \
  and\ \bibinfo {author} {\bibfnamefont {H.-H.}\ \bibnamefont {Tu}},\
  }\bibfield  {title} {\enquote {\bibinfo {title} {{SU}(3) trimer
  resonating-valence-bond state on the square lattice},}\ }\href {\doibase
  10.1103/PhysRevB.98.205117} {\bibfield  {journal} {\bibinfo  {journal} {Phys.
  Rev. B}\ }\textbf {\bibinfo {volume} {98}},\ \bibinfo {pages} {205117}
  (\bibinfo {year} {2018})}\BibitemShut {NoStop}%
\bibitem [{\citenamefont {Kim}\ \emph {et~al.}(2017)\citenamefont {Kim},
  \citenamefont {Penc}, \citenamefont {Nataf},\ and\ \citenamefont
  {Mila}}]{kim_linear_2017}%
  \BibitemOpen
  \bibfield  {author} {\bibinfo {author} {\bibfnamefont {F.~H.}\ \bibnamefont
  {Kim}}, \bibinfo {author} {\bibfnamefont {K.}~\bibnamefont {Penc}}, \bibinfo
  {author} {\bibfnamefont {P.}~\bibnamefont {Nataf}}, \ and\ \bibinfo {author}
  {\bibfnamefont {F.}~\bibnamefont {Mila}},\ }\bibfield  {title} {\enquote
  {\bibinfo {title} {Linear flavor-wave theory for fully antisymmetric
  {SU}(${N}$) irreducible representations},}\ }\href {\doibase
  10.1103/PhysRevB.96.205142} {\bibfield  {journal} {\bibinfo  {journal} {Phys.
  Rev. B}\ }\textbf {\bibinfo {volume} {96}},\ \bibinfo {pages} {205142}
  (\bibinfo {year} {2017})}\BibitemShut {NoStop}%
\bibitem [{\citenamefont {St\'ephan}\ \emph {et~al.}(2013)\citenamefont
  {St\'ephan}, \citenamefont {Ju}, \citenamefont {Fendley},\ and\ \citenamefont
  {Melko}}]{stephan_entanglement_2013}%
  \BibitemOpen
  \bibfield  {author} {\bibinfo {author} {\bibfnamefont {J.-M.}\ \bibnamefont
  {St\'ephan}}, \bibinfo {author} {\bibfnamefont {H.}~\bibnamefont {Ju}},
  \bibinfo {author} {\bibfnamefont {P.}~\bibnamefont {Fendley}}, \ and\
  \bibinfo {author} {\bibfnamefont {R.~G.}\ \bibnamefont {Melko}},\ }\bibfield
  {title} {\enquote {\bibinfo {title} {Entanglement in gapless
  resonating-valence-bond states},}\ }\href {\doibase
  10.1088/1367-2630/15/1/015004} {\bibfield  {journal} {\bibinfo  {journal}
  {New Journal of Physics}\ }\textbf {\bibinfo {volume} {15}},\ \bibinfo
  {pages} {015004} (\bibinfo {year} {2013})}\BibitemShut {NoStop}%
\bibitem [{sup()}]{supp}%
  \BibitemOpen
  \href@noop {} {}\bibinfo {howpublished} {See the supplementary material for
  more information.}\BibitemShut {Stop}%
\bibitem [{\citenamefont {Jiang}\ and\ \citenamefont
  {Ran}(2015)}]{jiang_symmetric_2015}%
  \BibitemOpen
  \bibfield  {author} {\bibinfo {author} {\bibfnamefont {S.}~\bibnamefont
  {Jiang}}\ and\ \bibinfo {author} {\bibfnamefont {Y.}~\bibnamefont {Ran}},\
  }\bibfield  {title} {\enquote {\bibinfo {title} {Symmetric tensor networks
  and practical simulation algorithms to sharply identify classes of quantum
  phases distinguishable by short-range physics},}\ }\href {\doibase
  10.1103/PhysRevB.92.104414} {\bibfield  {journal} {\bibinfo  {journal} {Phys.
  Rev. B}\ }\textbf {\bibinfo {volume} {92}},\ \bibinfo {pages} {104414}
  (\bibinfo {year} {2015})}\BibitemShut {NoStop}%
\bibitem [{\citenamefont {Or\'us}\ and\ \citenamefont
  {Vidal}(2009)}]{orus_simulation_2009}%
  \BibitemOpen
  \bibfield  {author} {\bibinfo {author} {\bibfnamefont {R.}~\bibnamefont
  {Or\'us}}\ and\ \bibinfo {author} {\bibfnamefont {G.}~\bibnamefont {Vidal}},\
  }\bibfield  {title} {\enquote {\bibinfo {title} {Simulation of
  two-dimensional quantum systems on an infinite lattice revisited: {Corner}
  transfer matrix for tensor contraction},}\ }\href {\doibase
  10.1103/PhysRevB.80.094403} {\bibfield  {journal} {\bibinfo  {journal} {Phys.
  Rev. B}\ }\textbf {\bibinfo {volume} {80}},\ \bibinfo {pages} {094403}
  (\bibinfo {year} {2009})}\BibitemShut {NoStop}%
\bibitem [{\citenamefont {Rokhsar}\ and\ \citenamefont
  {Kivelson}(1988)}]{rokhsar_superconductivity_1988}%
  \BibitemOpen
  \bibfield  {author} {\bibinfo {author} {\bibfnamefont {D.~S.}\ \bibnamefont
  {Rokhsar}}\ and\ \bibinfo {author} {\bibfnamefont {Steven~A.}\ \bibnamefont
  {Kivelson}},\ }\bibfield  {title} {\enquote {\bibinfo {title}
  {Superconductivity and the quantum hard--core dimer gas},}\ }\href {\doibase
  10.1103/PhysRevLett.61.2376} {\bibfield  {journal} {\bibinfo  {journal}
  {Phys. Rev. Lett.}\ }\textbf {\bibinfo {volume} {61}},\ \bibinfo {pages}
  {2376} (\bibinfo {year} {1988})}\BibitemShut {NoStop}%
\bibitem [{\citenamefont {Fradkin}\ \emph {et~al.}(2004)\citenamefont
  {Fradkin}, \citenamefont {Huse}, \citenamefont {Moessner}, \citenamefont
  {Oganesyan},\ and\ \citenamefont {Sondhi}}]{fradkin_rk_2004}%
  \BibitemOpen
  \bibfield  {author} {\bibinfo {author} {\bibfnamefont {E.}~\bibnamefont
  {Fradkin}}, \bibinfo {author} {\bibfnamefont {D.~A.}\ \bibnamefont {Huse}},
  \bibinfo {author} {\bibfnamefont {R.}~\bibnamefont {Moessner}}, \bibinfo
  {author} {\bibfnamefont {V.}~\bibnamefont {Oganesyan}}, \ and\ \bibinfo
  {author} {\bibfnamefont {S.~L.}\ \bibnamefont {Sondhi}},\ }\bibfield  {title}
  {\enquote {\bibinfo {title} {Bipartite {R}okhsar--{K}ivelson points and
  {C}antor deconfinement},}\ }\href {\doibase 10.1103/PhysRevB.69.224415}
  {\bibfield  {journal} {\bibinfo  {journal} {Phys. Rev. B}\ }\textbf {\bibinfo
  {volume} {69}},\ \bibinfo {pages} {224415} (\bibinfo {year}
  {2004})}\BibitemShut {NoStop}%
\bibitem [{\citenamefont {Damle}\ \emph {et~al.}(2012)\citenamefont {Damle},
  \citenamefont {Dhar},\ and\ \citenamefont {Ramola}}]{damle_resonating_2012}%
  \BibitemOpen
  \bibfield  {author} {\bibinfo {author} {\bibfnamefont {K.}~\bibnamefont
  {Damle}}, \bibinfo {author} {\bibfnamefont {D.}~\bibnamefont {Dhar}}, \ and\
  \bibinfo {author} {\bibfnamefont {K.}~\bibnamefont {Ramola}},\ }\bibfield
  {title} {\enquote {\bibinfo {title} {Resonating {Valence} {Bond} wave
  functions and classical interacting dimer models},}\ }\href {\doibase
  10.1103/PhysRevLett.108.247216} {\bibfield  {journal} {\bibinfo  {journal}
  {Phys. Rev. Lett.}\ }\textbf {\bibinfo {volume} {108}},\ \bibinfo {pages}
  {247216} (\bibinfo {year} {2012})}\BibitemShut {NoStop}%
\bibitem [{\citenamefont {Alet}\ \emph {et~al.}(2006)\citenamefont {Alet},
  \citenamefont {Ikhlef}, \citenamefont {Jacobsen}, \citenamefont {Misguich},\
  and\ \citenamefont {Pasquier}}]{alet_classical_2006}%
  \BibitemOpen
  \bibfield  {author} {\bibinfo {author} {\bibfnamefont {F.}~\bibnamefont
  {Alet}}, \bibinfo {author} {\bibfnamefont {Y.}~\bibnamefont {Ikhlef}},
  \bibinfo {author} {\bibfnamefont {J.~L.}\ \bibnamefont {Jacobsen}}, \bibinfo
  {author} {\bibfnamefont {G.}~\bibnamefont {Misguich}}, \ and\ \bibinfo
  {author} {\bibfnamefont {V.}~\bibnamefont {Pasquier}},\ }\bibfield  {title}
  {\enquote {\bibinfo {title} {Classical dimers with aligning interactions on
  the square lattice},}\ }\href {\doibase 10.1103/PhysRevE.74.041124}
  {\bibfield  {journal} {\bibinfo  {journal} {Phys. Rev. E}\ }\textbf {\bibinfo
  {volume} {74}},\ \bibinfo {pages} {041124} (\bibinfo {year}
  {2006})}\BibitemShut {NoStop}%
\bibitem [{\citenamefont {Calabrese}\ and\ \citenamefont
  {Cardy}(2004)}]{calabrese_entanglement_2004}%
  \BibitemOpen
  \bibfield  {author} {\bibinfo {author} {\bibfnamefont {P.}~\bibnamefont
  {Calabrese}}\ and\ \bibinfo {author} {\bibfnamefont {J.}~\bibnamefont
  {Cardy}},\ }\bibfield  {title} {\enquote {\bibinfo {title} {Entanglement
  entropy and quantum field theory},}\ }\href {\doibase
  10.1088/1742-5468/2004/06/P06002} {\bibfield  {journal} {\bibinfo  {journal}
  {J. Stat. Mech. Theory Exp.}\ }\textbf {\bibinfo {volume} {2004}},\ \bibinfo
  {pages} {P06002} (\bibinfo {year} {2004})}\BibitemShut {NoStop}%
\bibitem [{\citenamefont {Cirac}\ \emph {et~al.}(2011)\citenamefont {Cirac},
  \citenamefont {Poilblanc}, \citenamefont {Schuch},\ and\ \citenamefont
  {Verstraete}}]{cirac_entanglement_2011}%
  \BibitemOpen
  \bibfield  {author} {\bibinfo {author} {\bibfnamefont {J.~I.}\ \bibnamefont
  {Cirac}}, \bibinfo {author} {\bibfnamefont {D.}~\bibnamefont {Poilblanc}},
  \bibinfo {author} {\bibfnamefont {N.}~\bibnamefont {Schuch}}, \ and\ \bibinfo
  {author} {\bibfnamefont {F.}~\bibnamefont {Verstraete}},\ }\bibfield  {title}
  {\enquote {\bibinfo {title} {Entanglement spectrum and boundary theories with
  projected entangled-pair states},}\ }\href {\doibase
  10.1103/PhysRevB.83.245134} {\bibfield  {journal} {\bibinfo  {journal} {Phys.
  Rev. B}\ }\textbf {\bibinfo {volume} {83}},\ \bibinfo {pages} {245134}
  (\bibinfo {year} {2011})}\BibitemShut {NoStop}%
\bibitem [{\citenamefont {Jiang}\ \emph {et~al.}(2013)\citenamefont {Jiang},
  \citenamefont {Singh},\ and\ \citenamefont {Balents}}]{jiang_accuracy_2013}%
  \BibitemOpen
  \bibfield  {author} {\bibinfo {author} {\bibfnamefont {H.-C.}\ \bibnamefont
  {Jiang}}, \bibinfo {author} {\bibfnamefont {R.~R.~P.}\ \bibnamefont {Singh}},
  \ and\ \bibinfo {author} {\bibfnamefont {L.}~\bibnamefont {Balents}},\
  }\bibfield  {title} {\enquote {\bibinfo {title} {Accuracy of topological
  entanglement entropy on finite cylinders},}\ }\href {\doibase
  10.1103/PhysRevLett.111.107205} {\bibfield  {journal} {\bibinfo  {journal}
  {Phys. Rev. Lett.}\ }\textbf {\bibinfo {volume} {111}},\ \bibinfo {pages}
  {107205} (\bibinfo {year} {2013})}\BibitemShut {NoStop}%
\bibitem [{\citenamefont {He}\ \emph {et~al.}(2014)\citenamefont {He},
  \citenamefont {Moradi},\ and\ \citenamefont {Wen}}]{he_modular_2014}%
  \BibitemOpen
  \bibfield  {author} {\bibinfo {author} {\bibfnamefont {H.}~\bibnamefont
  {He}}, \bibinfo {author} {\bibfnamefont {H.}~\bibnamefont {Moradi}}, \ and\
  \bibinfo {author} {\bibfnamefont {X.-G.}\ \bibnamefont {Wen}},\ }\bibfield
  {title} {\enquote {\bibinfo {title} {Modular matrices as topological order
  parameter by a gauge-symmetry-preserved tensor renormalization approach},}\
  }\href {\doibase 10.1103/PhysRevB.90.205114} {\bibfield  {journal} {\bibinfo
  {journal} {Phys. Rev. B}\ }\textbf {\bibinfo {volume} {90}},\ \bibinfo
  {pages} {205114} (\bibinfo {year} {2014})}\BibitemShut {NoStop}%
\bibitem [{\citenamefont {Mei}\ \emph {et~al.}(2017)\citenamefont {Mei},
  \citenamefont {Chen}, \citenamefont {He},\ and\ \citenamefont
  {Wen}}]{mei_gapped_2017}%
  \BibitemOpen
  \bibfield  {author} {\bibinfo {author} {\bibfnamefont {J.-W.}\ \bibnamefont
  {Mei}}, \bibinfo {author} {\bibfnamefont {J.-Y.}\ \bibnamefont {Chen}},
  \bibinfo {author} {\bibfnamefont {H.}~\bibnamefont {He}}, \ and\ \bibinfo
  {author} {\bibfnamefont {X.-G.}\ \bibnamefont {Wen}},\ }\bibfield  {title}
  {\enquote {\bibinfo {title} {Gapped spin liquid with $\mathbb{Z}_2$
  topological order for the kagome {Heisenberg} model},}\ }\href {\doibase
  10.1103/PhysRevB.95.235107} {\bibfield  {journal} {\bibinfo  {journal} {Phys.
  Rev. B}\ }\textbf {\bibinfo {volume} {95}},\ \bibinfo {pages} {235107}
  (\bibinfo {year} {2017})}\BibitemShut {NoStop}%
\bibitem [{\citenamefont {Oshikawa}(2000)}]{Oshikawa2000}%
  \BibitemOpen
  \bibfield  {author} {\bibinfo {author} {\bibfnamefont {M.}~\bibnamefont
  {Oshikawa}},\ }\bibfield  {title} {\enquote {\bibinfo {title}
  {Commensurability, excitation gap, and topology in quantum many-particle
  systems on a periodic lattice},}\ }\href {\doibase
  10.1103/PhysRevLett.84.1535} {\bibfield  {journal} {\bibinfo  {journal}
  {Phys. Rev. Lett.}\ }\textbf {\bibinfo {volume} {84}},\ \bibinfo {pages}
  {1535} (\bibinfo {year} {2000})}\BibitemShut {NoStop}%
\bibitem [{\citenamefont {Lieb}\ \emph {et~al.}(1961)\citenamefont {Lieb},
  \citenamefont {Schultz},\ and\ \citenamefont {Mattis}}]{Lieb1961}%
  \BibitemOpen
  \bibfield  {author} {\bibinfo {author} {\bibfnamefont {E.}~\bibnamefont
  {Lieb}}, \bibinfo {author} {\bibfnamefont {T.}~\bibnamefont {Schultz}}, \
  and\ \bibinfo {author} {\bibfnamefont {D.}~\bibnamefont {Mattis}},\
  }\bibfield  {title} {\enquote {\bibinfo {title} {Two soluble models of an
  antiferromagnetic chain},}\ }\href {\doibase
  https://doi.org/10.1016/0003-4916(61)90115-4} {\bibfield  {journal} {\bibinfo
   {journal} {Annals of Physics}\ }\textbf {\bibinfo {volume} {16}},\ \bibinfo
  {pages} {407} (\bibinfo {year} {1961})}\BibitemShut {NoStop}%
\bibitem [{\citenamefont {Hastings}(2004)}]{Hastings2004}%
  \BibitemOpen
  \bibfield  {author} {\bibinfo {author} {\bibfnamefont {M.~B.}\ \bibnamefont
  {Hastings}},\ }\bibfield  {title} {\enquote {\bibinfo {title}
  {{L}ieb-{S}chultz-{M}attis in higher dimensions},}\ }\href {\doibase
  10.1103/PhysRevB.69.104431} {\bibfield  {journal} {\bibinfo  {journal} {Phys.
  Rev. B}\ }\textbf {\bibinfo {volume} {69}},\ \bibinfo {pages} {104431}
  (\bibinfo {year} {2004})}\BibitemShut {NoStop}%
\bibitem [{Note1()}]{Note1}%
  \BibitemOpen
  \bibinfo {note} {K. Totsuka, private communication. Similarly, a
  non-degenerate featureless gapped ground state is not possible on the square
  lattice with SU(4) fundamental and antifundamental representation on the two
  sublattices~\cite {Jian2018}.}\BibitemShut {Stop}%
\bibitem [{\citenamefont {Paramekanti}\ and\ \citenamefont
  {Marston}(2007)}]{paramekanti2007}%
  \BibitemOpen
  \bibfield  {author} {\bibinfo {author} {\bibfnamefont {A.}~\bibnamefont
  {Paramekanti}}\ and\ \bibinfo {author} {\bibfnamefont {J.~B.}\ \bibnamefont
  {Marston}},\ }\bibfield  {title} {\enquote {\bibinfo {title} {{SU(N)} quantum
  spin models: a variational wavefunction study},}\ }\href
  {https://doi.org/10.1088/0953-8984/19/12/125215} {\bibfield  {journal}
  {\bibinfo  {journal} {J. Phys.: Condens. Matter}\ }\textbf {\bibinfo {volume}
  {19}},\ \bibinfo {pages} {125215} (\bibinfo {year} {2007})}\BibitemShut
  {NoStop}%
\bibitem [{\citenamefont {Wang}\ \emph {et~al.}(2014)\citenamefont {Wang},
  \citenamefont {Li}, \citenamefont {Cai}, \citenamefont {Zhou}, \citenamefont
  {Wang},\ and\ \citenamefont {Wu}}]{wang2014}%
  \BibitemOpen
  \bibfield  {author} {\bibinfo {author} {\bibfnamefont {D.}~\bibnamefont
  {Wang}}, \bibinfo {author} {\bibfnamefont {Y.}~\bibnamefont {Li}}, \bibinfo
  {author} {\bibfnamefont {Z.}~\bibnamefont {Cai}}, \bibinfo {author}
  {\bibfnamefont {Z.}~\bibnamefont {Zhou}}, \bibinfo {author} {\bibfnamefont
  {Y.}~\bibnamefont {Wang}}, \ and\ \bibinfo {author} {\bibfnamefont
  {C.}~\bibnamefont {Wu}},\ }\bibfield  {title} {\enquote {\bibinfo {title}
  {Competing orders in the {2D} half-filled {SU(2N)} {H}ubbard model through
  the pinning-field quantum {Monte Carlo} simulations},}\ }\href {\doibase
  10.1103/PhysRevLett.112.156403} {\bibfield  {journal} {\bibinfo  {journal}
  {Phys. Rev. Lett.}\ }\textbf {\bibinfo {volume} {112}},\ \bibinfo {pages}
  {156403} (\bibinfo {year} {2014})}\BibitemShut {NoStop}%
\bibitem [{\citenamefont {Boos}\ \emph {et~al.}(2018)\citenamefont {Boos},
  \citenamefont {Lajk\'o}, \citenamefont {Nataf}, \citenamefont {Penc},
  \citenamefont {Schmidt},\ and\ \citenamefont {Mila}}]{boos_chiral_2018}%
  \BibitemOpen
  \bibfield  {author} {\bibinfo {author} {\bibfnamefont {C.}~\bibnamefont
  {Boos}}, \bibinfo {author} {\bibfnamefont {M.}~\bibnamefont {Lajk\'o}},
  \bibinfo {author} {\bibfnamefont {P.}~\bibnamefont {Nataf}}, \bibinfo
  {author} {\bibfnamefont {K.}~\bibnamefont {Penc}}, \bibinfo {author}
  {\bibfnamefont {K.~P.}\ \bibnamefont {Schmidt}}, \ and\ \bibinfo {author}
  {\bibfnamefont {F.}~\bibnamefont {Mila}},\ }\bibfield  {title} {\enquote
  {\bibinfo {title} {Chiral {Mott} phase of three-component fermions on the
  triangular lattice},}\ }\href {http://arxiv.org/abs/1802.03179} {\bibfield
  {journal} {\bibinfo  {journal} {arXiv:1802.03179 [cond-mat]}\ } (\bibinfo
  {year} {2018})}\BibitemShut {NoStop}%
\bibitem [{\citenamefont {Pagano}\ \emph {et~al.}(2014)\citenamefont {Pagano},
  \citenamefont {Mancini}, \citenamefont {Cappellini}, \citenamefont
  {Lombardi}, \citenamefont {Sch\"afer}, \citenamefont {Hu}, \citenamefont
  {Liu}, \citenamefont {Catani}, \citenamefont {Sias}, \citenamefont
  {Inguscio},\ and\ \citenamefont {Fallani}}]{Pagano2014}%
  \BibitemOpen
  \bibfield  {author} {\bibinfo {author} {\bibfnamefont {G.}~\bibnamefont
  {Pagano}}, \bibinfo {author} {\bibfnamefont {M.}~\bibnamefont {Mancini}},
  \bibinfo {author} {\bibfnamefont {G.}~\bibnamefont {Cappellini}}, \bibinfo
  {author} {\bibfnamefont {P.}~\bibnamefont {Lombardi}}, \bibinfo {author}
  {\bibfnamefont {F.}~\bibnamefont {Sch\"afer}}, \bibinfo {author}
  {\bibfnamefont {H.}~\bibnamefont {Hu}}, \bibinfo {author} {\bibfnamefont
  {X.-J.}\ \bibnamefont {Liu}}, \bibinfo {author} {\bibfnamefont
  {J.}~\bibnamefont {Catani}}, \bibinfo {author} {\bibfnamefont
  {C.}~\bibnamefont {Sias}}, \bibinfo {author} {\bibfnamefont {M.}~\bibnamefont
  {Inguscio}}, \ and\ \bibinfo {author} {\bibfnamefont {L.}~\bibnamefont
  {Fallani}},\ }\bibfield  {title} {\enquote {\bibinfo {title} {A
  one-dimensional liquid of fermions with tunable spin},}\ }\href
  {https://doi.org/10.1038/nphys2878} {\bibfield  {journal} {\bibinfo
  {journal} {Nature Physics}\ }\textbf {\bibinfo {volume} {10}},\ \bibinfo
  {pages} {198} (\bibinfo {year} {2014})}\BibitemShut {NoStop}%
\bibitem [{\citenamefont {Jian}\ \emph {et~al.}(2018)\citenamefont {Jian},
  \citenamefont {Bi},\ and\ \citenamefont {Xu}}]{Jian2018}%
  \BibitemOpen
  \bibfield  {author} {\bibinfo {author} {\bibfnamefont {C.-M.}\ \bibnamefont
  {Jian}}, \bibinfo {author} {\bibfnamefont {Z.}~\bibnamefont {Bi}}, \ and\
  \bibinfo {author} {\bibfnamefont {C.}~\bibnamefont {Xu}},\ }\bibfield
  {title} {\enquote {\bibinfo {title} {{Lieb}-{Schultz}-{Mattis} theorem and
  its generalizations from the perspective of the symmetry-protected
  topological phase},}\ }\href {\doibase 10.1103/PhysRevB.97.054412} {\bibfield
   {journal} {\bibinfo  {journal} {Phys. Rev. B}\ }\textbf {\bibinfo {volume}
  {97}},\ \bibinfo {pages} {054412} (\bibinfo {year} {2018})}\BibitemShut
  {NoStop}%
\end{thebibliography}%


\begin{thebibliography}{4}%
\makeatletter
\providecommand \@ifxundefined [1]{%
 \@ifx{#1\undefined}
}%
\providecommand \@ifnum [1]{%
 \ifnum #1\expandafter \@firstoftwo
 \else \expandafter \@secondoftwo
 \fi
}%
\providecommand \@ifx [1]{%
 \ifx #1\expandafter \@firstoftwo
 \else \expandafter \@secondoftwo
 \fi
}%
\providecommand \natexlab [1]{#1}%
\providecommand \enquote  [1]{``#1''}%
\providecommand \bibnamefont  [1]{#1}%
\providecommand \bibfnamefont [1]{#1}%
\providecommand \citenamefont [1]{#1}%
\providecommand \href@noop [0]{\@secondoftwo}%
\providecommand \href [0]{\begingroup \@sanitize@url \@href}%
\providecommand \@href[1]{\@@startlink{#1}\@@href}%
\providecommand \@@href[1]{\endgroup#1\@@endlink}%
\providecommand \@sanitize@url [0]{\catcode `\\12\catcode `\$12\catcode
  `\&12\catcode `\#12\catcode `\^12\catcode `\_12\catcode `\%12\relax}%
\providecommand \@@startlink[1]{}%
\providecommand \@@endlink[0]{}%
\providecommand \url  [0]{\begingroup\@sanitize@url \@url }%
\providecommand \@url [1]{\endgroup\@href {#1}{\urlprefix }}%
\providecommand \urlprefix  [0]{URL }%
\providecommand \Eprint [0]{\href }%
\providecommand \doibase [0]{http://dx.doi.org/}%
\providecommand \selectlanguage [0]{\@gobble}%
\providecommand \bibinfo  [0]{\@secondoftwo}%
\providecommand \bibfield  [0]{\@secondoftwo}%
\providecommand \translation [1]{[#1]}%
\providecommand \BibitemOpen [0]{}%
\providecommand \bibitemStop [0]{}%
\providecommand \bibitemNoStop [0]{.\EOS\space}%
\providecommand \EOS [0]{\spacefactor3000\relax}%
\providecommand \BibitemShut  [1]{\csname bibitem#1\endcsname}%
\let\auto@bib@innerbib\@empty
\bibitem [{\citenamefont {Or\'us}\ and\ \citenamefont
  {Vidal}(2009)}]{orus_simulation_2009}%
  \BibitemOpen
  \bibfield  {author} {\bibinfo {author} {\bibfnamefont {R.}~\bibnamefont
  {Or\'us}}\ and\ \bibinfo {author} {\bibfnamefont {G.}~\bibnamefont {Vidal}},\
  }\bibfield  {title} {\enquote {\bibinfo {title} {Simulation of
  two-dimensional quantum systems on an infinite lattice revisited: {Corner}
  transfer matrix for tensor contraction},}\ }\href {\doibase
  10.1103/PhysRevB.80.094403} {\bibfield  {journal} {\bibinfo  {journal} {Phys.
  Rev. B}\ }\textbf {\bibinfo {volume} {80}},\ \bibinfo {pages} {094403}
  (\bibinfo {year} {2009})}\BibitemShut {NoStop}%
\bibitem [{\citenamefont {Cirac}\ \emph {et~al.}(2011)\citenamefont {Cirac},
  \citenamefont {Poilblanc}, \citenamefont {Schuch},\ and\ \citenamefont
  {Verstraete}}]{cirac_entanglement_2011}%
  \BibitemOpen
  \bibfield  {author} {\bibinfo {author} {\bibfnamefont {J.~I.}\ \bibnamefont
  {Cirac}}, \bibinfo {author} {\bibfnamefont {D.}~\bibnamefont {Poilblanc}},
  \bibinfo {author} {\bibfnamefont {N.}~\bibnamefont {Schuch}}, \ and\ \bibinfo
  {author} {\bibfnamefont {F.}~\bibnamefont {Verstraete}},\ }\bibfield  {title}
  {\enquote {\bibinfo {title} {Entanglement spectrum and boundary theories with
  projected entangled-pair states},}\ }\href {\doibase
  10.1103/PhysRevB.83.245134} {\bibfield  {journal} {\bibinfo  {journal} {Phys.
  Rev. B}\ }\textbf {\bibinfo {volume} {83}},\ \bibinfo {pages} {245134}
  (\bibinfo {year} {2011})}\BibitemShut {NoStop}%
\bibitem [{\citenamefont {Jiang}\ \emph {et~al.}(2013)\citenamefont {Jiang},
  \citenamefont {Singh},\ and\ \citenamefont {Balents}}]{jiang_accuracy_2013}%
  \BibitemOpen
  \bibfield  {author} {\bibinfo {author} {\bibfnamefont {H.-C.}\ \bibnamefont
  {Jiang}}, \bibinfo {author} {\bibfnamefont {R.~R.~P.}\ \bibnamefont {Singh}},
  \ and\ \bibinfo {author} {\bibfnamefont {L.}~\bibnamefont {Balents}},\
  }\bibfield  {title} {\enquote {\bibinfo {title} {Accuracy of topological
  entanglement entropy on finite cylinders},}\ }\href {\doibase
  10.1103/PhysRevLett.111.107205} {\bibfield  {journal} {\bibinfo  {journal}
  {Phys. Rev. Lett.}\ }\textbf {\bibinfo {volume} {111}},\ \bibinfo {pages}
  {107205} (\bibinfo {year} {2013})}\BibitemShut {NoStop}%
\bibitem [{\citenamefont {Li}\ and\ \citenamefont
  {Haldane}(2008)}]{li_entanglement_2008}%
  \BibitemOpen
  \bibfield  {author} {\bibinfo {author} {\bibfnamefont {H.}~\bibnamefont
  {Li}}\ and\ \bibinfo {author} {\bibfnamefont {F.~D.~M.}\ \bibnamefont
  {Haldane}},\ }\bibfield  {title} {\enquote {\bibinfo {title} {Entanglement
  {Spectrum} as a {Generalization} of {Entanglement} {Entropy}:
  {Identification} of {Topological} {Order} in {Non}-{Abelian} {Fractional}
  {Quantum} {Hall} {Effect} {States}},}\ }\href {\doibase
  10.1103/PhysRevLett.101.010504} {\bibfield  {journal} {\bibinfo  {journal}
  {Physical Review Letters}\ }\textbf {\bibinfo {volume} {101}},\ \bibinfo
  {pages} {010504} (\bibinfo {year} {2008})}\BibitemShut {NoStop}%
\end{thebibliography}%



\end{document}




\title{
Supplemental Materials: $SU(4)$ topological RVB spin liquid on the square lattice
}
	

%
\author{Olivier Gauth\'e}
\affiliation{Laboratoire de Physique Th\'eorique, Universit\'e de Toulouse, CNRS, UPS, France}
\author{Sylvain Capponi}
\affiliation{Laboratoire de Physique Th\'eorique, Universit\'e de Toulouse, CNRS, UPS, France}
\author{Didier Poilblanc}
\affiliation{Laboratoire de Physique Th\'eorique, Universit\'e de Toulouse, CNRS, UPS, France}
\affiliation{Institute of Theoretical Physics, \'Ecole Polytechnique F\'ed\'erale de Lausanne (EPFL), CH-1015 Lausanne, Switzerland}


\date{\today}

\maketitle

\section{Bond operator}
\
Two singlets can be made from the two representations $\textbf{6}$ or $\textbf{1}$ of SU(4). Thus, there exist two linearly independent projectors $(\textbf{6}\oplus\textbf{1})^{\otimes 2}\rightarrow \textbf{1}$
on a singlet state, which are matrices of dimension $(49,1)$. We reshape them as matrices $(7,7)$ (which are no more projectors in the mathematical sense) and as explained in the article we choose an angle $\pi/4$ between the projectors $\mathcal{P}_{\textbf{6}\otimes \textbf{6}} \otimes \mathbb{1}$ and $\mathbb{1} \otimes \mathcal{P}_{\textbf{1}\otimes \textbf{1}}$. 
The states of the  $\textbf{6}$-representation labelled as $|0\big>$, $|1\big>$, $|2\big>$, $|3\big>$, $|4\big>$, $|5\big>$  are defined by their (Cartan) quantum numbers $(1, 0, -1)$, $(1, -1, 1)$, $(0, -1, 0)$, $(0, 1, 0)$, $(-1, 1, -1)$ and $(-1, 0, 1)$, respectively. The last $|6\big>$ state corresponds to the $\textbf{1}$-singlet.
With this convention  of ordering the vectors
of the $\textbf{6}\oplus \textbf{1}$ representation, the projector we apply reads
\begin{equation}
P = 1/\sqrt{12}\begin{pmatrix}
0 & 0 & 0 & 0 & 0 & 1 & 0\\
0 & 0 & 0 & 0 & -1 & 0 & 0\\
0 & 0 & 0 & 1 & 0 & 0 & 0\\	
0 & 0 & 1 & 0 & 0 & 0 & 0\\
0 & -1 & 0 & 0 & 0 & 0 & 0\\
1 & 0 & 0 & 0 & 0 & 0 & 0\\
0 & 0 & 0 & 0 & 0 & 0 & \sqrt{6}\\
\end{pmatrix}.
\end{equation}

To avoid dealing explicitly with it in our tensor network algorithm, we absorb $P$ in the definition of the tensor $A$. More precisely, we consider the square root of $P$ - which is a complex symmetric matrix - and contract it on every physical leg of the initial tensor $A$ as shown in Fig \ref{fig:CTM_full} $(a)$. The double layer tensor 
$\mathbb{E}$ is then computed after this operation is done and it turns out that it also exhibits the same $A_1$+i$A_2$ symmetry as the original $A$ tensor. 

\section{CTMRG algorithm}

\begin{figure}
	\centering
		\includegraphics[width=0.45\textwidth]{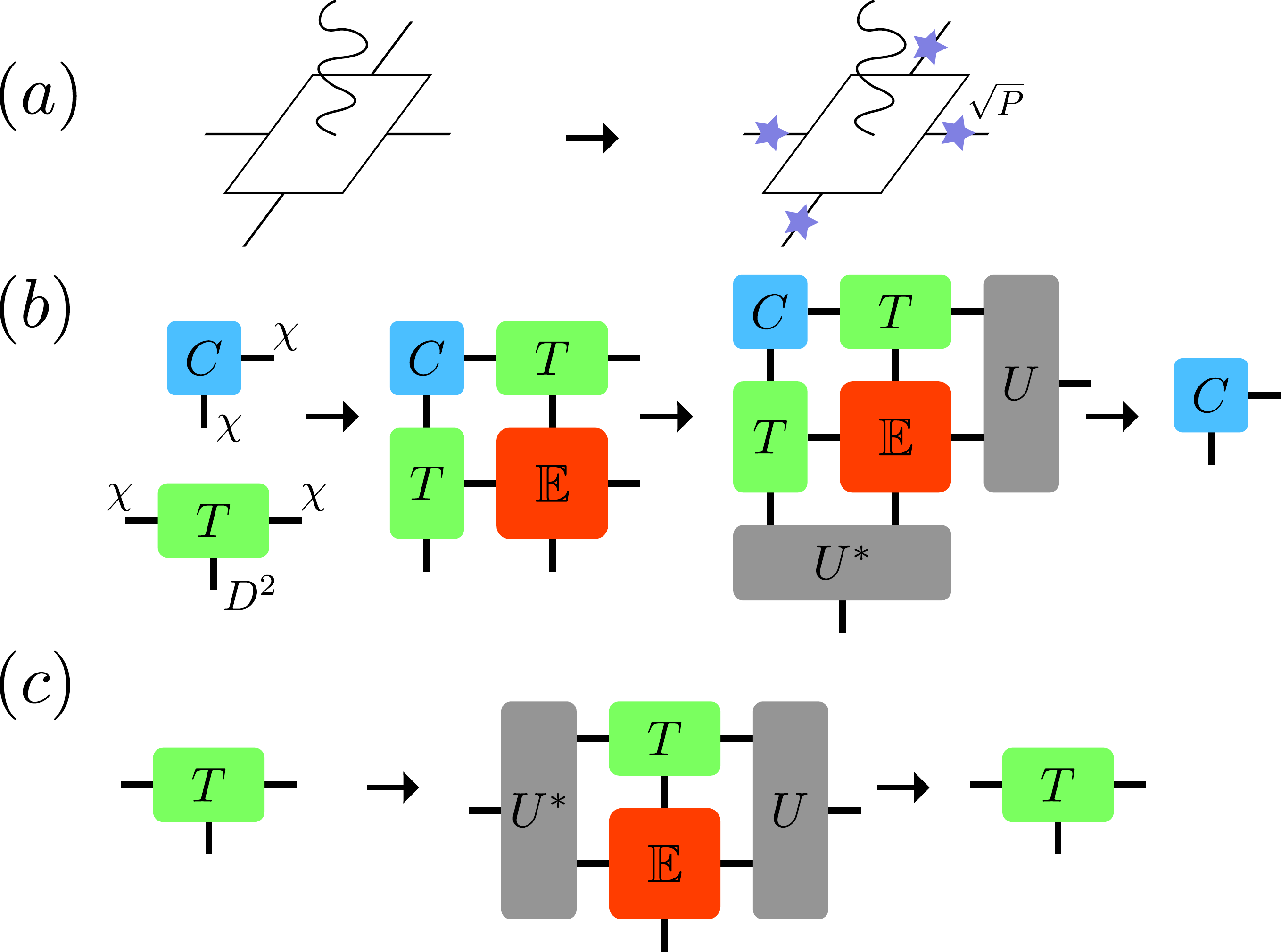}
		\caption{\footnotesize{($a$) The square root of the projector on the singlet $\sqrt{P}$ is added to every virtual leg of the initial tensor $A$. We will then omit this matrix for clarity. Corner Transfer Matrix Renormalization Group (CTMRG) algorithm~\cite{orus_simulation_2009}: ($b$) Renormalization of the corner matrix $C$. ($c$) Renormalization of the side tensor $T$.}}
		\label{fig:CTM_full}
\end{figure}

Thanks to rotation invariance, we renormalize only one corner $C$ instead of doing a full update on the four corners. While diagonalizing the new corner (see Fig. \ref{fig:CTM_full} $(b)$), we obtain a unitary transfer matrix $U$. This matrix is used to renormalize the edge tensor $T$, where we add another tensor $\mathbb{E}$ to $T$ as shown in Fig. \ref{fig:CTM_full} $(c)$.

\section{Correlation lengths and correlation functions} 
\begin{figure}
  \centering
    \includegraphics[width=0.4\textwidth]{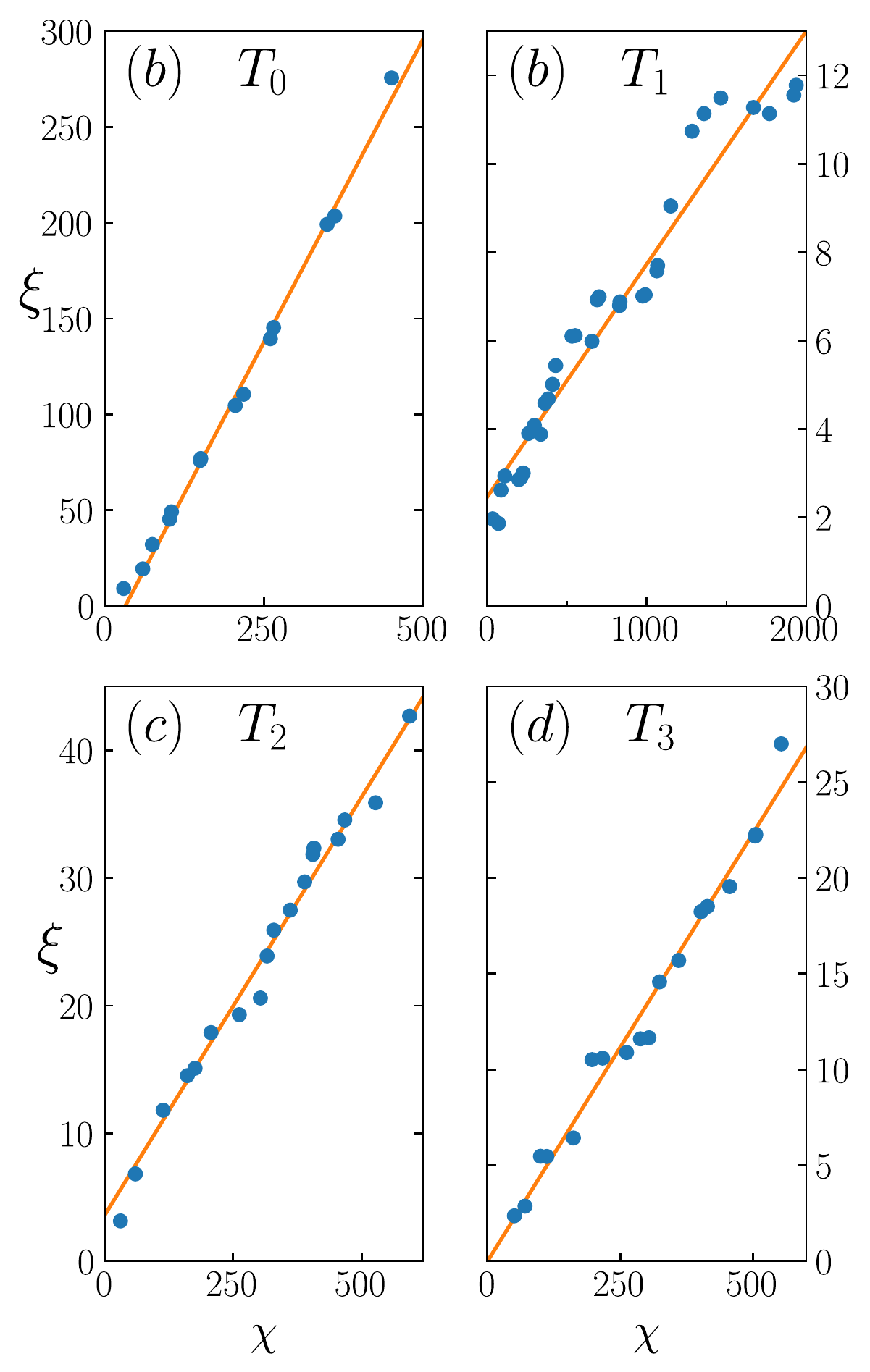}
    \caption{\footnotesize{The maximum correlation length $\xi$ of the system is plotted for the four PEPS given by the elementary tensors $T_0$, $T_1$, $T_2$ and $T_3$, as a function of the cut parameter $\chi$ of the CTMRG algorithm. Fits are linear, which attest that the wavefunctions are critical.}}
    \label{fig:corr1T}
\end{figure}

We have computed the (maximum) correlation length $\xi$ of the system using the Lanczos algorithm  to diagonalize the approximate transfer matrix. $\xi$ is a function of the corner matrix dimension $\chi$ and for a gapped wavefunction, the correlation length converges exponentially to a finite value $\xi_\infty$. For a critical wavefunction, $\xi$ grows linearly with $\chi$ as we can see in Fig. \ref{fig:corr1T} for the wavefunctions defined by $T_0$, $T_1$, $T_2$ and $T_3$. 
Note that, in that case, the maximum correlation length extracted from the spectrum of the transfer matrix corresponds in fact to the diverging dimer correlation length (see below).
Also, for $T_1$, the slope of $\xi$ vs $\chi$ is quite small so that only moderate $\xi\sim 6$ or $7$ are accessible even for $\chi\sim 700$.  This renders the estimation of the (2+0) central charge and dimer critical exponent more difficult
for $T_1$.

\begin{figure}
	\centering
		\includegraphics[width=0.4\textwidth]{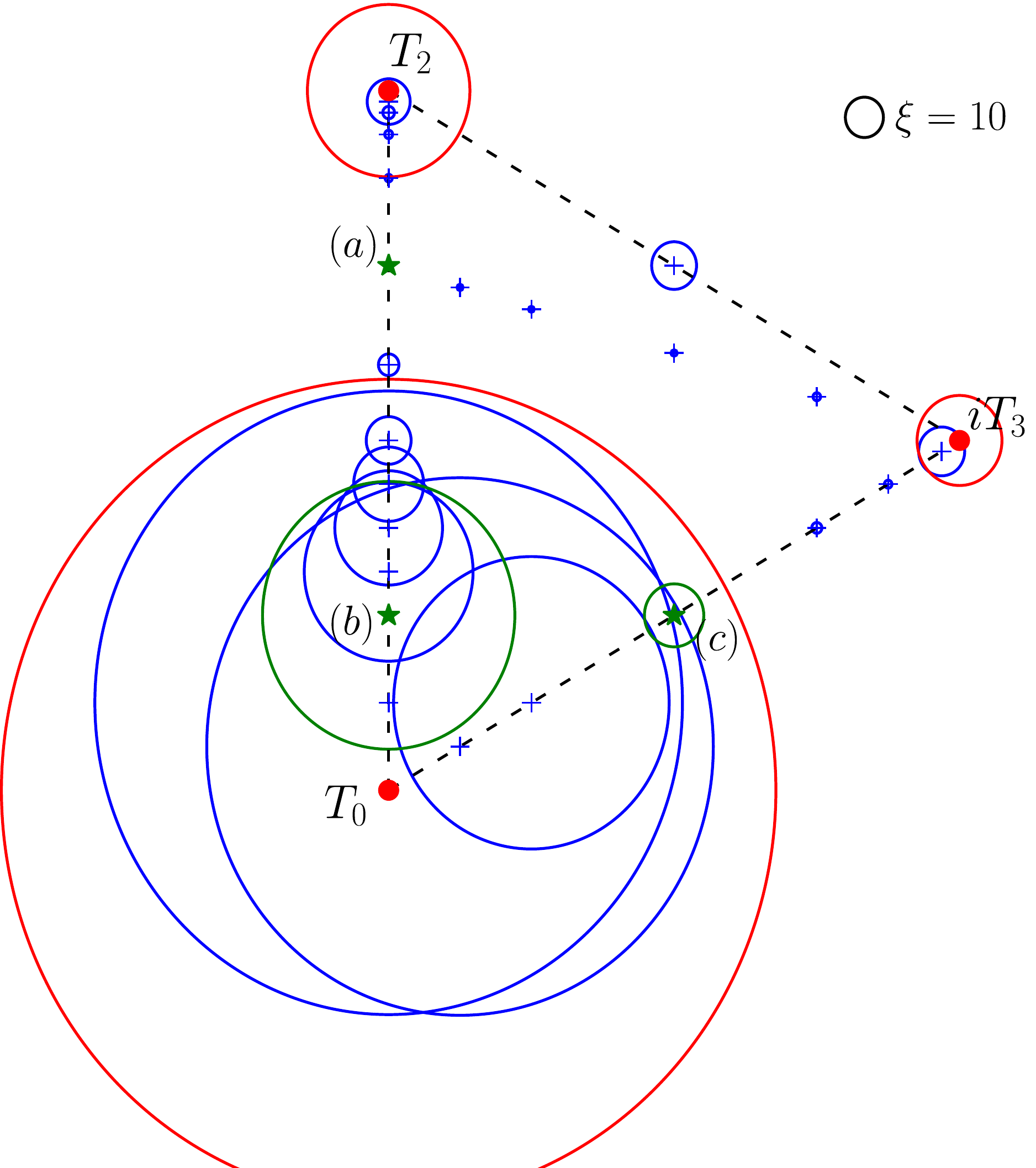}
		\caption{\footnotesize{We consider the (restricted) two-dimensional tensor family given by 
		$A = \cos \theta \cos \phi \, T_2 + \sin \theta \cos \phi \, T_0 + i \sin \phi \, T_3$ and we plot the maximal value of $\xi$ we obtained (the value of $\chi$ is not exactly the same for every circle). The red points label the three elementary tensors, the green stars label the points where the boundary entanglement entropy and the entanglement spectrum were computed (see figures (\ref{fig:sR_pannel}) and (\ref{fig:spectra})).}}
		\label{fig:circles}
\end{figure}

Letting aside the tensor $T_1$, we have a 2-parameters space parametrized by
\begin{equation}
A(\theta,\phi) = \cos\theta\cos\phi \, T_2 + \sin\theta\cos\phi \, T_0 + i\sin\phi \, T_3
\end{equation}
We show the values of the (maximum) correlation length in this space in Fig \ref{fig:circles} for several values of $\theta$ and $\phi$. The cut parameter $\chi$ was taken around 350 but may be slightly different for the different circles. We note that $\xi$ gets larger when $A$ is approaching one of the $T_0$, $T_2$ or $T_3$ corners of the triangular parameter space. It is however unclear on this plot whether there exists any critical region in the vicinity of the three elementary critical wavefunctions. Nevertheless, a finite-$\chi$ scaling analysis suggests that there is indeed such a (small) extended critical 
region around the $T_0$ corner (see main text). 

\begin{figure}
  \centering
    \includegraphics[width=0.4\textwidth]{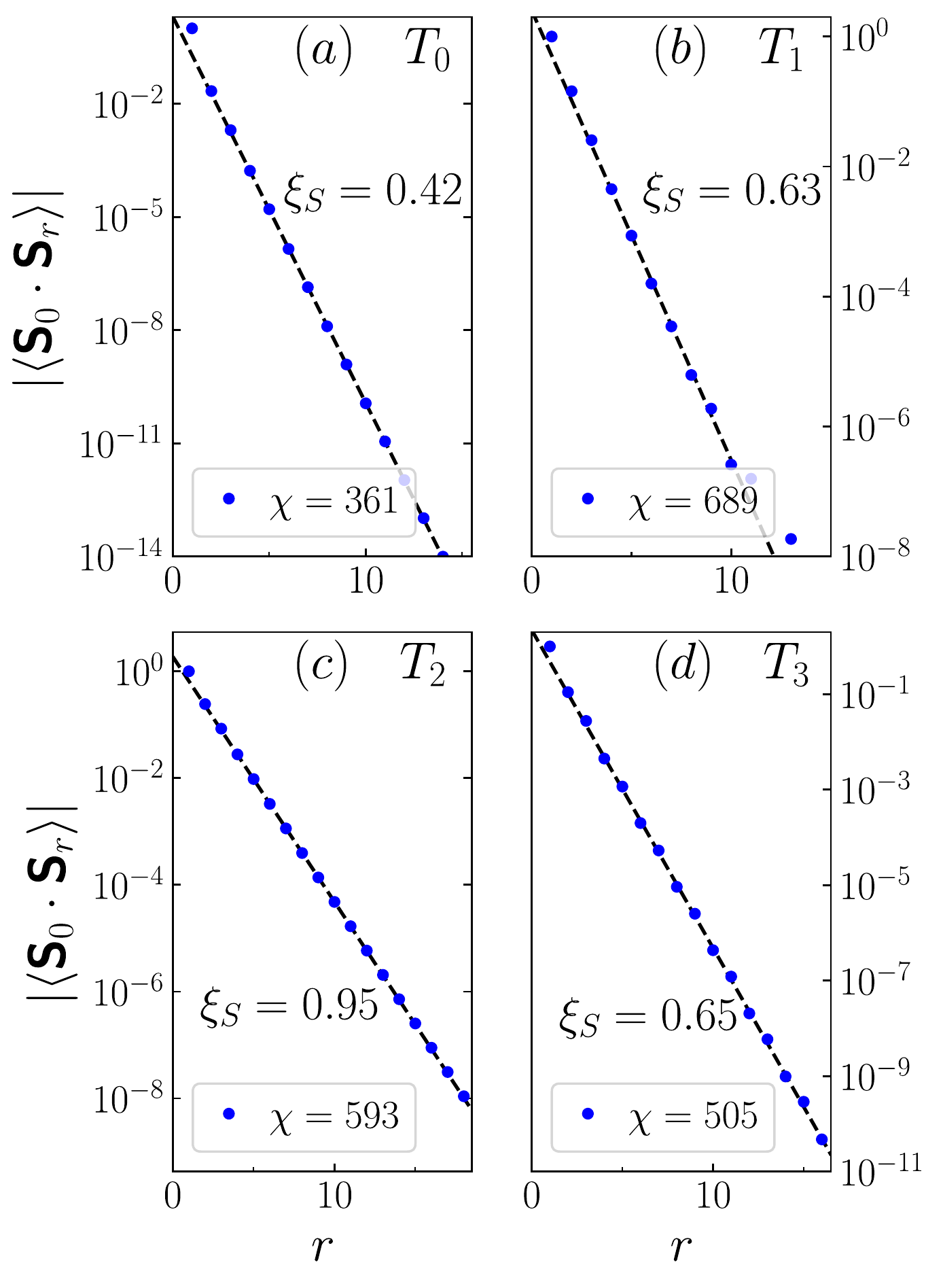}
    \caption{\footnotesize{The "spin" correlation functions are plotted for the four elementary PEPS associated to the $T_0$, $T_1$, $T_2$, and $T_3$ tensors. We observe an exponential decay with a "spin" correlation length $\xi_S$.}}
    \label{figSz_1T}
\end{figure}

\begin{figure}
  \centering
    \includegraphics[width=0.4\textwidth]{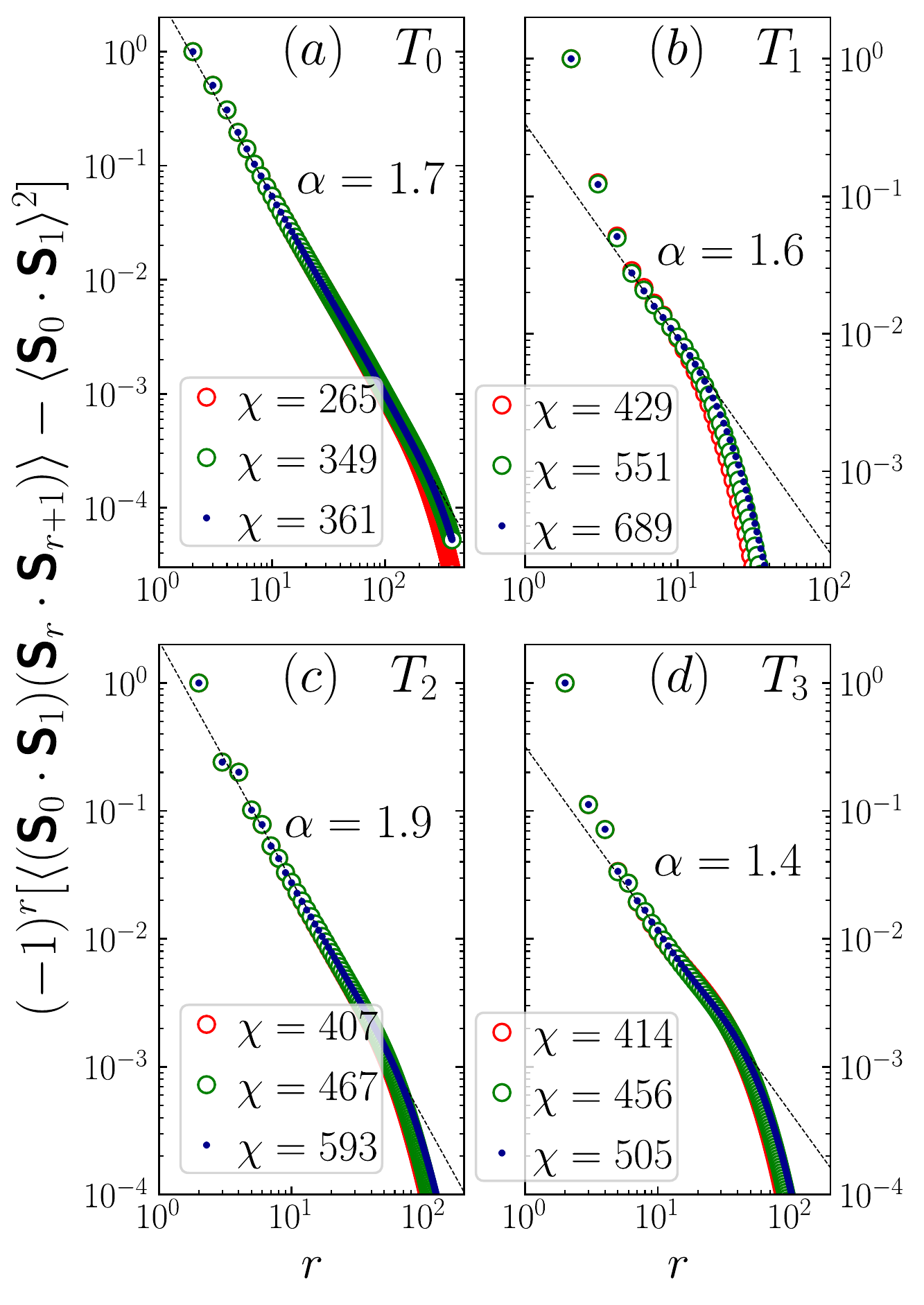}
    \caption{\footnotesize{The dimer-dimer correlation functions are plotted for the four elementary tensors $T_0$, $T_1$, $T_2$, and $T_3$. We observe a critical behavior with algebraic decay which is cut at large distances by an exponential fall-off. The range of the algebraic behavior extends as $\chi$ grows (and would become infinite for
    $\chi\rightarrow\infty$).}}
    \label{fig:dimers1T}
\end{figure}

From the converged $C$ and $T$ tensors one can construct the environment of any rectangular subsystem. Using an (infinitely long)  strip 
delimited by two chains of $T$ tensors,
 we have computed first the expectation value of the observable $\textbf{S}_i\cdot \textbf{S}_j$ for all distance $|i-j|$ in the strip direction, as plotted in Fig \ref{figSz_1T} for the four tensors $T_i$.  We observe a clear exponential decay with a very short "spin" correlation length $\xi_S$. We have also computed the dimer-dimer correlation function in Fig \ref{fig:dimers1T}. This observable has long range correlations with algebraic decay, below a finite-$\chi$ induced length scale (of the order of the maximum correlation length
 $\xi(\chi)$). As a consequence, only for large enough $\chi$ can one fit the algebraic behavior on a sufficiently large range of distances to obtain accurate
 values of the critical exponent.

\section{Boundary properties}

Let us consider the PEPS on an infinitely long cylinder of finite circumference $N_v$ partitioned into two halves. $N_v$ copies of the tensor $T$ are contracted to construct the edges of the two semi-infinite half-cylinders. This vector of dimension  
$D^{2N v}$ is then reshaped into a boundary matrix $\sigma_{b}$ of shape $D^{N v} \times D^{N_v} $. The operator $\rho_{b} = \sigma_{b}^2$ defined on the one-dimensional virtual space of the cut is related by an isometry (which maps two-dimensions to one-dimension and conserves the spectrum) to the actual reduced 
density matrix~\cite{cirac_entanglement_2011} defined by tracing out the physical degrees of freedom of half of the infinite cylinder. The operator $\rho_b$ splits into two $\mathbb{Z}_2$ topological sectors and we normalize those blocks independently to have both traces equal to 1.

\begin{figure}
	\centering
		\includegraphics[width=0.45\textwidth]{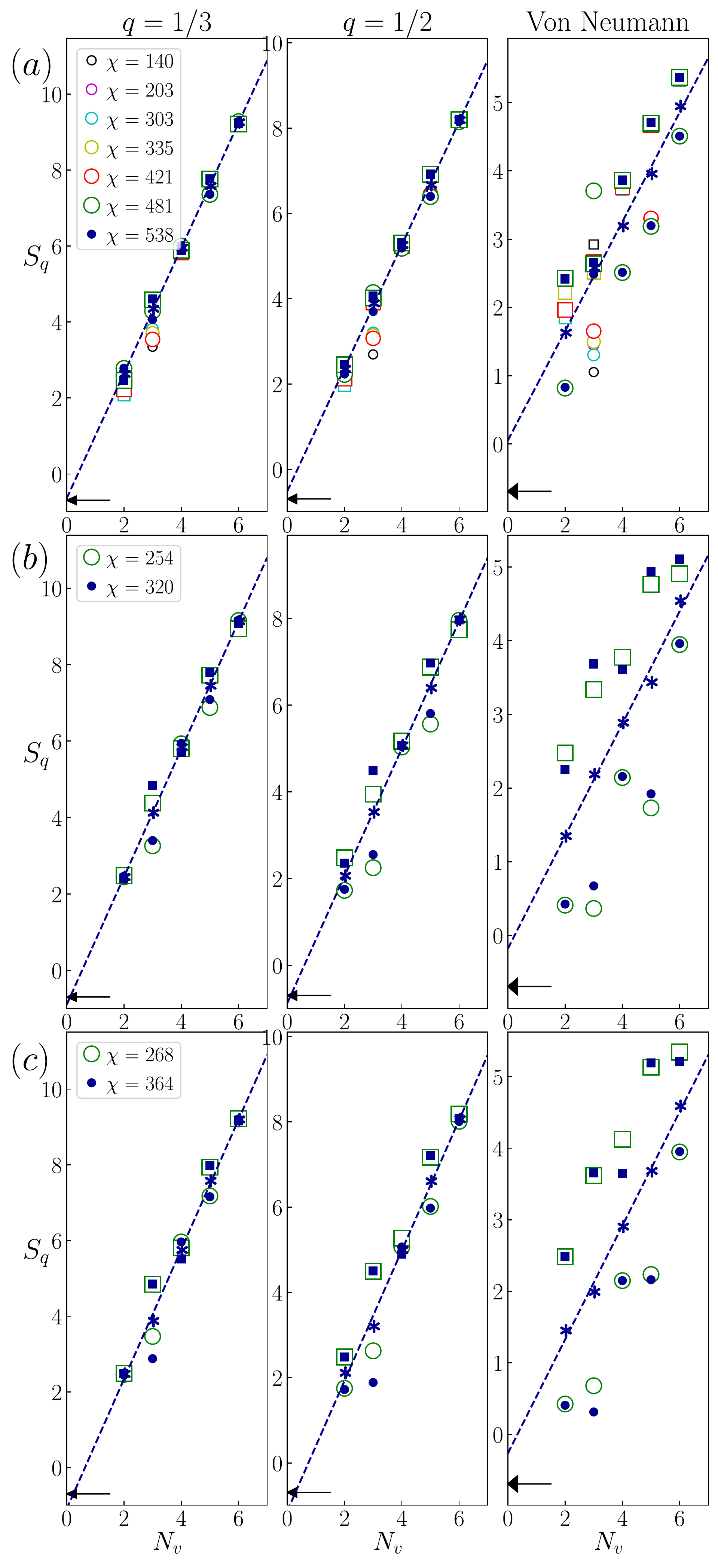}
		\caption{\footnotesize{Renyi entanglement entropies of a partitioned cylinder for the wavefunctions defined in Fig.~\ref{fig:circles}, with respect to the cylinder circumference $N_v$ for $q = 1/3, 1/2$ and 1. The squares label the odd sector and the circles the even one. The stars stand for the average of the two sectors and the dashed lines are their linear fits, the arrow points to the value $-\log2$.}}
		\label{fig:sR_pannel}
\end{figure}

The Renyi entanglement entropy associated to the partition of the cylinder is defined from the boundary density matrix $\rho_{b}$ as $S_q = \log \Tr \rho_{b}^q /(1-q)$, the Von Neumann entropy being the limit $q \rightarrow 1$. 
The entanglement entropy satisfies the area low i.e. it scales with the length of the cut $N_v$ and its sub-leading correction (constant topological entropy) characterizes the topological nature of the SL.
We have split the entropy into its two contributions coming from the even and odd topological sectors (note that the definitions of even and odd are exchanged when $N_v$ is odd) in order to observe the $-\log 2$ topological entropy expected for a $\mathbb{Z}_2$ SL.
We have computed $S_q$ for three values of the parameter $\theta$ and $\phi$, labeled in Fig \ref{fig:circles}: $(a)$ $A = \cos (\pi/8) T_2 + \sin (\pi/8) T_0$, $(b)$ $A = \cos (3\pi/8) T_2 + \sin (3\pi/8) T_0$, $(c)$ $A = \cos (\pi/4) T_0 + i \sin (\pi/4) T_3$ and plotted it for $q=1/3, 1/2$ and $1$ in Fig \ref{fig:sR_pannel}. We note that as $q$ grows, the extrapolation on the vertical axis deviates substantially from the expected $-\log 2$ value, in agreement with Ref.~\onlinecite{jiang_accuracy_2013}.

\begin{figure}
	\centering
		\includegraphics[width=0.45\textwidth]{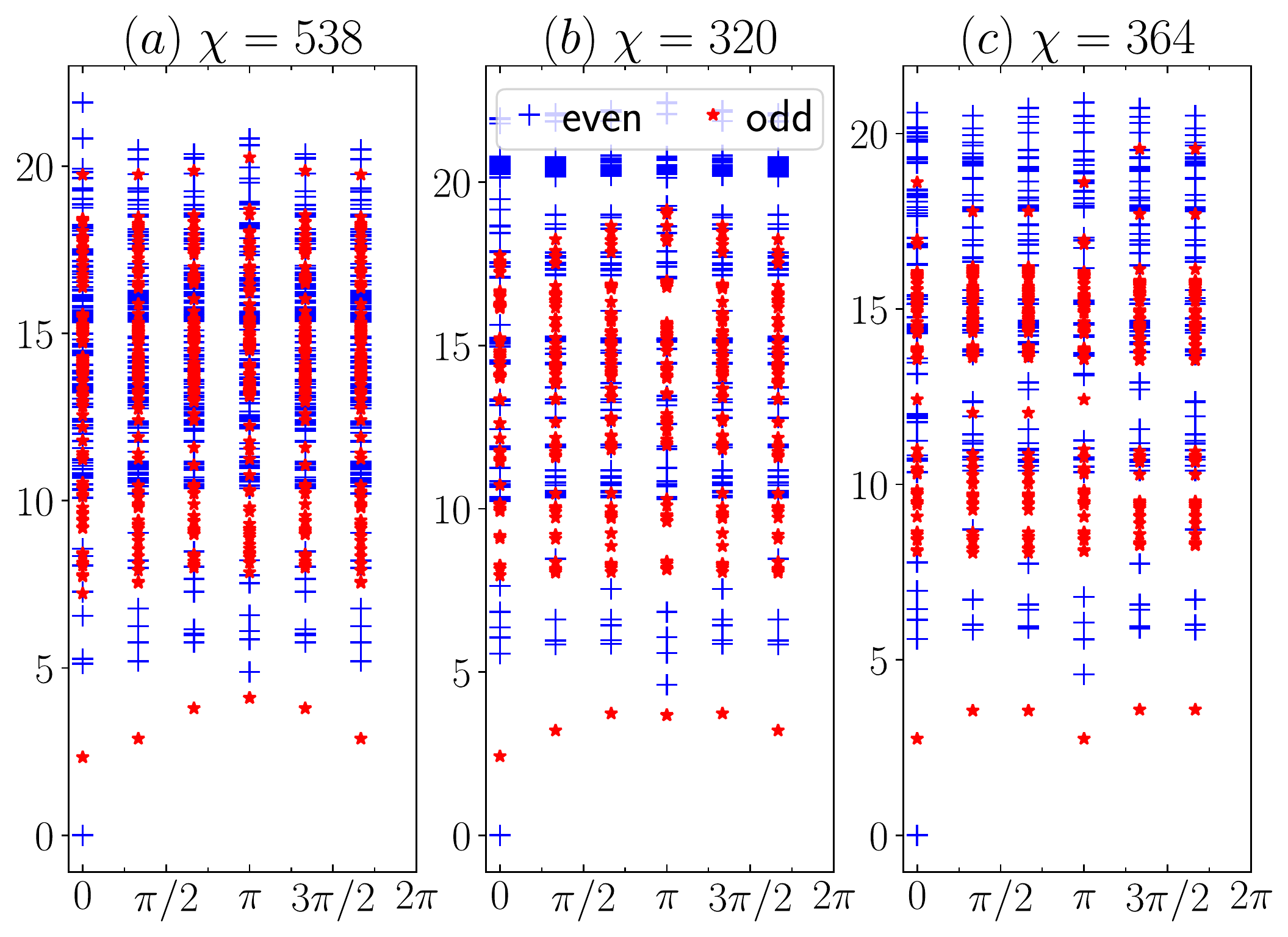}
		\caption{\footnotesize{Entanglement spectrum of a semi-infinite cylinder of circumference $N_v = 6$ with respect to the wavevector along the edge in the two topological even and odd sectors ($a$) $A = \cos (\pi/8) T_2 + \sin (\pi/8) T_0$ with $\chi = 303$ ($b$) $A = \cos (3\pi/8) T_2 + \sin (3\pi/8) T_0$ with $\chi = 254$ ($c$) $A = \cos (\pi/4) T_0 + i \sin (\pi/4) T_3$ with $\chi = 268$ (see Fig. (\ref{fig:circles}))} for the location in the parameter space.}
		\label{fig:spectra}
\end{figure}

The entanglement spectrum~\cite{li_entanglement_2008} is the spectrum of the entanglement Hamiltonian $\mathcal{H}$ defined as $\rho_b = \exp(-\mathcal{H})$.  The entanglement spectrum is shown in Fig \ref{fig:spectra} for the three points $(a)$, $(b)$ and $(c)$. The ground state of $\mathcal{H}$ is always an even singlet and the first excitation is an odd $\textbf{6}$ irrep. 
We believe the excitation energy remains finite for $N_v\rightarrow\infty$ i.e. the entanglement spectrum remains {\it gapped}. 
The points $(a)$ and $(b)$ correspond to real wavefunctions and we indeed observe a $k \longleftrightarrow -k$ symmetry in the spectrum
associated to time-reversal ($\cal T$) and parity ($\cal P$) symmetries. The point $(c)$ labels a complex wavefunction with $A_1 + i A_2$ symmetry that breaks $\cal T$ and $\cal P$. We observe that the even sector of the entanglement spectrum still exhibits 
the  $k \longleftrightarrow -k$ symmetry, but not the odd sector, which follows a $k \longleftrightarrow \pi-k$ symmetry. We believe this is due to the fact that the
product $\cal TP$ is still preserved. Note also that we have not observed the emergence of 
gapless chiral edge modes, the entanglement spectrum being probably gapped. 

\section{Tensor expressions}

For clarity, we give here the coefficients of the \emph{unnormalized} tensors $T_i$ which are integer values. However, the parametrization of the $A$ PEPS tensor involves the $T_i$ tensors normalized with the Frobenius norm. These four tensors are orthogonal for the dot product defined by this norm. We recall that the first index labels the physical variable (varying from 0 to 5) and the four subsequent indices the virtual 
variables (varying from 0 to 6) on the links (in e.g. clockwise direction).


\begin{longtable}{MMM}
\hline \hline
T_0[0,0,6,6,6] = 1 & T_0[0,6,0,6,6] = 1 & T_0[0,6,6,0,6] = 1 \\ 
T_0[0,6,6,6,0] = 1 & T_0[1,1,6,6,6] = 1 & T_0[1,6,1,6,6] = 1 \\ 
T_0[1,6,6,1,6] = 1 & T_0[1,6,6,6,1] = 1 & T_0[2,2,6,6,6] = 1 \\ 
T_0[2,6,2,6,6] = 1 & T_0[2,6,6,2,6] = 1 & T_0[2,6,6,6,2] = 1 \\ 
T_0[3,3,6,6,6] = 1 & T_0[3,6,3,6,6] = 1 & T_0[3,6,6,3,6] = 1 \\ 
T_0[3,6,6,6,3] = 1 & T_0[4,4,6,6,6] = 1 & T_0[4,6,4,6,6] = 1 \\ 
T_0[4,6,6,4,6] = 1 & T_0[4,6,6,6,4] = 1 & T_0[5,5,6,6,6] = 1 \\ 
T_0[5,6,5,6,6] = 1 & T_0[5,6,6,5,6] = 1 & T_0[5,6,6,6,5] = 1 \\
\hline \hline
\caption{Tensor $T_0$ (multiplied by $2 \sqrt{6}$).}
\label{Table:T0}
\end{longtable}


\begin{longtable}{MMM}
\hline \hline
T_1[0,0,0,5,6] = 2 & T_1[0,0,0,6,5] = 2 & T_1[0,0,1,4,6] = -1 \\ 
T_1[0,0,1,6,4] = -1 & T_1[0,0,2,3,6] = 1 & T_1[0,0,2,6,3] = 1 \\ 
T_1[0,0,3,2,6] = 1 & T_1[0,0,3,6,2] = 1 & T_1[0,0,4,1,6] = -1 \\ 
T_1[0,0,4,6,1] = -1 & T_1[0,0,5,0,6] = 2 & T_1[0,0,5,6,0] = 2 \\ 
T_1[0,0,6,0,5] = 2 & T_1[0,0,6,1,4] = -1 & T_1[0,0,6,2,3] = 1 \\ 
T_1[0,0,6,3,2] = 1 & T_1[0,0,6,4,1] = -1 & T_1[0,0,6,5,0] = 2 \\ 
T_1[0,1,0,4,6] = -1 & T_1[0,1,0,6,4] = -1 & T_1[0,1,4,0,6] = -1 \\ 
T_1[0,1,4,6,0] = -1 & T_1[0,1,6,0,4] = -1 & T_1[0,1,6,4,0] = -1 \\ 
T_1[0,2,0,3,6] = 1 & T_1[0,2,0,6,3] = 1 & T_1[0,2,3,0,6] = 1 \\ 
T_1[0,2,3,6,0] = 1 & T_1[0,2,6,0,3] = 1 & T_1[0,2,6,3,0] = 1 \\ 
T_1[0,3,0,2,6] = 1 & T_1[0,3,0,6,2] = 1 & T_1[0,3,2,0,6] = 1 \\ 
T_1[0,3,2,6,0] = 1 & T_1[0,3,6,0,2] = 1 & T_1[0,3,6,2,0] = 1 \\ 
T_1[0,4,0,1,6] = -1 & T_1[0,4,0,6,1] = -1 & T_1[0,4,1,0,6] = -1 \\ 
T_1[0,4,1,6,0] = -1 & T_1[0,4,6,0,1] = -1 & T_1[0,4,6,1,0] = -1 \\ 
T_1[0,5,0,0,6] = 2 & T_1[0,5,0,6,0] = 2 & T_1[0,5,6,0,0] = 2 \\ 
T_1[0,6,0,0,5] = 2 & T_1[0,6,0,1,4] = -1 & T_1[0,6,0,2,3] = 1 \\ 
T_1[0,6,0,3,2] = 1 & T_1[0,6,0,4,1] = -1 & T_1[0,6,0,5,0] = 2 \\ 
T_1[0,6,1,0,4] = -1 & T_1[0,6,1,4,0] = -1 & T_1[0,6,2,0,3] = 1 \\ 
T_1[0,6,2,3,0] = 1 & T_1[0,6,3,0,2] = 1 & T_1[0,6,3,2,0] = 1 \\ 
T_1[0,6,4,0,1] = -1 & T_1[0,6,4,1,0] = -1 & T_1[0,6,5,0,0] = 2 \\ 
T_1[1,0,1,5,6] = 1 & T_1[1,0,1,6,5] = 1 & T_1[1,0,5,1,6] = 1 \\ 
T_1[1,0,5,6,1] = 1 & T_1[1,0,6,1,5] = 1 & T_1[1,0,6,5,1] = 1 \\ 
T_1[1,1,0,5,6] = 1 & T_1[1,1,0,6,5] = 1 & T_1[1,1,1,4,6] = -2 \\ 
T_1[1,1,1,6,4] = -2 & T_1[1,1,2,3,6] = 1 & T_1[1,1,2,6,3] = 1 \\ 
T_1[1,1,3,2,6] = 1 & T_1[1,1,3,6,2] = 1 & T_1[1,1,4,1,6] = -2 \\ 
T_1[1,1,4,6,1] = -2 & T_1[1,1,5,0,6] = 1 & T_1[1,1,5,6,0] = 1 \\ 
T_1[1,1,6,0,5] = 1 & T_1[1,1,6,1,4] = -2 & T_1[1,1,6,2,3] = 1 \\ 
T_1[1,1,6,3,2] = 1 & T_1[1,1,6,4,1] = -2 & T_1[1,1,6,5,0] = 1 \\ 
T_1[1,2,1,3,6] = 1 & T_1[1,2,1,6,3] = 1 & T_1[1,2,3,1,6] = 1 \\ 
T_1[1,2,3,6,1] = 1 & T_1[1,2,6,1,3] = 1 & T_1[1,2,6,3,1] = 1 \\ 
T_1[1,3,1,2,6] = 1 & T_1[1,3,1,6,2] = 1 & T_1[1,3,2,1,6] = 1 \\ 
T_1[1,3,2,6,1] = 1 & T_1[1,3,6,1,2] = 1 & T_1[1,3,6,2,1] = 1 \\ 
T_1[1,4,1,1,6] = -2 & T_1[1,4,1,6,1] = -2 & T_1[1,4,6,1,1] = -2 \\ 
T_1[1,5,0,1,6] = 1 & T_1[1,5,0,6,1] = 1 & T_1[1,5,1,0,6] = 1 \\ 
T_1[1,5,1,6,0] = 1 & T_1[1,5,6,0,1] = 1 & T_1[1,5,6,1,0] = 1 \\ 
T_1[1,6,0,1,5] = 1 & T_1[1,6,0,5,1] = 1 & T_1[1,6,1,0,5] = 1 \\ 
T_1[1,6,1,1,4] = -2 & T_1[1,6,1,2,3] = 1 & T_1[1,6,1,3,2] = 1 \\ 
T_1[1,6,1,4,1] = -2 & T_1[1,6,1,5,0] = 1 & T_1[1,6,2,1,3] = 1 \\ 
T_1[1,6,2,3,1] = 1 & T_1[1,6,3,1,2] = 1 & T_1[1,6,3,2,1] = 1 \\ 
T_1[1,6,4,1,1] = -2 & T_1[1,6,5,0,1] = 1 & T_1[1,6,5,1,0] = 1 \\ 
T_1[2,0,2,5,6] = 1 & T_1[2,0,2,6,5] = 1 & T_1[2,0,5,2,6] = 1 \\ 
T_1[2,0,5,6,2] = 1 & T_1[2,0,6,2,5] = 1 & T_1[2,0,6,5,2] = 1 \\ 
T_1[2,1,2,4,6] = -1 & T_1[2,1,2,6,4] = -1 & T_1[2,1,4,2,6] = -1 \\ 
T_1[2,1,4,6,2] = -1 & T_1[2,1,6,2,4] = -1 & T_1[2,1,6,4,2] = -1 \\ 
T_1[2,2,0,5,6] = 1 & T_1[2,2,0,6,5] = 1 & T_1[2,2,1,4,6] = -1 \\ 
T_1[2,2,1,6,4] = -1 & T_1[2,2,2,3,6] = 2 & T_1[2,2,2,6,3] = 2 \\ 
T_1[2,2,3,2,6] = 2 & T_1[2,2,3,6,2] = 2 & T_1[2,2,4,1,6] = -1 \\ 
T_1[2,2,4,6,1] = -1 & T_1[2,2,5,0,6] = 1 & T_1[2,2,5,6,0] = 1 \\ 
T_1[2,2,6,0,5] = 1 & T_1[2,2,6,1,4] = -1 & T_1[2,2,6,2,3] = 2 \\ 
T_1[2,2,6,3,2] = 2 & T_1[2,2,6,4,1] = -1 & T_1[2,2,6,5,0] = 1 \\ 
T_1[2,3,2,2,6] = 2 & T_1[2,3,2,6,2] = 2 & T_1[2,3,6,2,2] = 2 \\ 
T_1[2,4,1,2,6] = -1 & T_1[2,4,1,6,2] = -1 & T_1[2,4,2,1,6] = -1 \\ 
T_1[2,4,2,6,1] = -1 & T_1[2,4,6,1,2] = -1 & T_1[2,4,6,2,1] = -1 \\ 
T_1[2,5,0,2,6] = 1 & T_1[2,5,0,6,2] = 1 & T_1[2,5,2,0,6] = 1 \\ 
T_1[2,5,2,6,0] = 1 & T_1[2,5,6,0,2] = 1 & T_1[2,5,6,2,0] = 1 \\ 
T_1[2,6,0,2,5] = 1 & T_1[2,6,0,5,2] = 1 & T_1[2,6,1,2,4] = -1 \\ 
T_1[2,6,1,4,2] = -1 & T_1[2,6,2,0,5] = 1 & T_1[2,6,2,1,4] = -1 \\ 
T_1[2,6,2,2,3] = 2 & T_1[2,6,2,3,2] = 2 & T_1[2,6,2,4,1] = -1 \\ 
T_1[2,6,2,5,0] = 1 & T_1[2,6,3,2,2] = 2 & T_1[2,6,4,1,2] = -1 \\ 
T_1[2,6,4,2,1] = -1 & T_1[2,6,5,0,2] = 1 & T_1[2,6,5,2,0] = 1 \\ 
T_1[3,0,3,5,6] = 1 & T_1[3,0,3,6,5] = 1 & T_1[3,0,5,3,6] = 1 \\ 
T_1[3,0,5,6,3] = 1 & T_1[3,0,6,3,5] = 1 & T_1[3,0,6,5,3] = 1 \\ 
T_1[3,1,3,4,6] = -1 & T_1[3,1,3,6,4] = -1 & T_1[3,1,4,3,6] = -1 \\ 
T_1[3,1,4,6,3] = -1 & T_1[3,1,6,3,4] = -1 & T_1[3,1,6,4,3] = -1 \\ 
T_1[3,2,3,3,6] = 2 & T_1[3,2,3,6,3] = 2 & T_1[3,2,6,3,3] = 2 \\ 
T_1[3,3,0,5,6] = 1 & T_1[3,3,0,6,5] = 1 & T_1[3,3,1,4,6] = -1 \\ 
T_1[3,3,1,6,4] = -1 & T_1[3,3,2,3,6] = 2 & T_1[3,3,2,6,3] = 2 \\ 
T_1[3,3,3,2,6] = 2 & T_1[3,3,3,6,2] = 2 & T_1[3,3,4,1,6] = -1 \\ 
T_1[3,3,4,6,1] = -1 & T_1[3,3,5,0,6] = 1 & T_1[3,3,5,6,0] = 1 \\ 
T_1[3,3,6,0,5] = 1 & T_1[3,3,6,1,4] = -1 & T_1[3,3,6,2,3] = 2 \\ 
T_1[3,3,6,3,2] = 2 & T_1[3,3,6,4,1] = -1 & T_1[3,3,6,5,0] = 1 \\ 
T_1[3,4,1,3,6] = -1 & T_1[3,4,1,6,3] = -1 & T_1[3,4,3,1,6] = -1 \\ 
T_1[3,4,3,6,1] = -1 & T_1[3,4,6,1,3] = -1 & T_1[3,4,6,3,1] = -1 \\ 
T_1[3,5,0,3,6] = 1 & T_1[3,5,0,6,3] = 1 & T_1[3,5,3,0,6] = 1 \\ 
T_1[3,5,3,6,0] = 1 & T_1[3,5,6,0,3] = 1 & T_1[3,5,6,3,0] = 1 \\ 
T_1[3,6,0,3,5] = 1 & T_1[3,6,0,5,3] = 1 & T_1[3,6,1,3,4] = -1 \\ 
T_1[3,6,1,4,3] = -1 & T_1[3,6,2,3,3] = 2 & T_1[3,6,3,0,5] = 1 \\ 
T_1[3,6,3,1,4] = -1 & T_1[3,6,3,2,3] = 2 & T_1[3,6,3,3,2] = 2 \\ 
T_1[3,6,3,4,1] = -1 & T_1[3,6,3,5,0] = 1 & T_1[3,6,4,1,3] = -1 \\ 
T_1[3,6,4,3,1] = -1 & T_1[3,6,5,0,3] = 1 & T_1[3,6,5,3,0] = 1 \\ 
T_1[4,0,4,5,6] = 1 & T_1[4,0,4,6,5] = 1 & T_1[4,0,5,4,6] = 1 \\ 
T_1[4,0,5,6,4] = 1 & T_1[4,0,6,4,5] = 1 & T_1[4,0,6,5,4] = 1 \\ 
T_1[4,1,4,4,6] = -2 & T_1[4,1,4,6,4] = -2 & T_1[4,1,6,4,4] = -2 \\ 
T_1[4,2,3,4,6] = 1 & T_1[4,2,3,6,4] = 1 & T_1[4,2,4,3,6] = 1 \\ 
T_1[4,2,4,6,3] = 1 & T_1[4,2,6,3,4] = 1 & T_1[4,2,6,4,3] = 1 \\ 
T_1[4,3,2,4,6] = 1 & T_1[4,3,2,6,4] = 1 & T_1[4,3,4,2,6] = 1 \\ 
T_1[4,3,4,6,2] = 1 & T_1[4,3,6,2,4] = 1 & T_1[4,3,6,4,2] = 1 \\ 
T_1[4,4,0,5,6] = 1 & T_1[4,4,0,6,5] = 1 & T_1[4,4,1,4,6] = -2 \\ 
T_1[4,4,1,6,4] = -2 & T_1[4,4,2,3,6] = 1 & T_1[4,4,2,6,3] = 1 \\ 
T_1[4,4,3,2,6] = 1 & T_1[4,4,3,6,2] = 1 & T_1[4,4,4,1,6] = -2 \\ 
T_1[4,4,4,6,1] = -2 & T_1[4,4,5,0,6] = 1 & T_1[4,4,5,6,0] = 1 \\ 
T_1[4,4,6,0,5] = 1 & T_1[4,4,6,1,4] = -2 & T_1[4,4,6,2,3] = 1 \\ 
T_1[4,4,6,3,2] = 1 & T_1[4,4,6,4,1] = -2 & T_1[4,4,6,5,0] = 1 \\ 
T_1[4,5,0,4,6] = 1 & T_1[4,5,0,6,4] = 1 & T_1[4,5,4,0,6] = 1 \\ 
T_1[4,5,4,6,0] = 1 & T_1[4,5,6,0,4] = 1 & T_1[4,5,6,4,0] = 1 \\ 
T_1[4,6,0,4,5] = 1 & T_1[4,6,0,5,4] = 1 & T_1[4,6,1,4,4] = -2 \\ 
T_1[4,6,2,3,4] = 1 & T_1[4,6,2,4,3] = 1 & T_1[4,6,3,2,4] = 1 \\ 
T_1[4,6,3,4,2] = 1 & T_1[4,6,4,0,5] = 1 & T_1[4,6,4,1,4] = -2 \\ 
T_1[4,6,4,2,3] = 1 & T_1[4,6,4,3,2] = 1 & T_1[4,6,4,4,1] = -2 \\ 
T_1[4,6,4,5,0] = 1 & T_1[4,6,5,0,4] = 1 & T_1[4,6,5,4,0] = 1 \\ 
T_1[5,0,5,5,6] = 2 & T_1[5,0,5,6,5] = 2 & T_1[5,0,6,5,5] = 2 \\ 
T_1[5,1,4,5,6] = -1 & T_1[5,1,4,6,5] = -1 & T_1[5,1,5,4,6] = -1 \\ 
T_1[5,1,5,6,4] = -1 & T_1[5,1,6,4,5] = -1 & T_1[5,1,6,5,4] = -1 \\ 
T_1[5,2,3,5,6] = 1 & T_1[5,2,3,6,5] = 1 & T_1[5,2,5,3,6] = 1 \\ 
T_1[5,2,5,6,3] = 1 & T_1[5,2,6,3,5] = 1 & T_1[5,2,6,5,3] = 1 \\ 
T_1[5,3,2,5,6] = 1 & T_1[5,3,2,6,5] = 1 & T_1[5,3,5,2,6] = 1 \\ 
T_1[5,3,5,6,2] = 1 & T_1[5,3,6,2,5] = 1 & T_1[5,3,6,5,2] = 1 \\ 
T_1[5,4,1,5,6] = -1 & T_1[5,4,1,6,5] = -1 & T_1[5,4,5,1,6] = -1 \\ 
T_1[5,4,5,6,1] = -1 & T_1[5,4,6,1,5] = -1 & T_1[5,4,6,5,1] = -1 \\ 
T_1[5,5,0,5,6] = 2 & T_1[5,5,0,6,5] = 2 & T_1[5,5,1,4,6] = -1 \\ 
T_1[5,5,1,6,4] = -1 & T_1[5,5,2,3,6] = 1 & T_1[5,5,2,6,3] = 1 \\ 
T_1[5,5,3,2,6] = 1 & T_1[5,5,3,6,2] = 1 & T_1[5,5,4,1,6] = -1 \\ 
T_1[5,5,4,6,1] = -1 & T_1[5,5,5,0,6] = 2 & T_1[5,5,5,6,0] = 2 \\ 
T_1[5,5,6,0,5] = 2 & T_1[5,5,6,1,4] = -1 & T_1[5,5,6,2,3] = 1 \\ 
T_1[5,5,6,3,2] = 1 & T_1[5,5,6,4,1] = -1 & T_1[5,5,6,5,0] = 2 \\ 
T_1[5,6,0,5,5] = 2 & T_1[5,6,1,4,5] = -1 & T_1[5,6,1,5,4] = -1 \\ 
T_1[5,6,2,3,5] = 1 & T_1[5,6,2,5,3] = 1 & T_1[5,6,3,2,5] = 1 \\ 
T_1[5,6,3,5,2] = 1 & T_1[5,6,4,1,5] = -1 & T_1[5,6,4,5,1] = -1 \\ 
T_1[5,6,5,0,5] = 2 & T_1[5,6,5,1,4] = -1 & T_1[5,6,5,2,3] = 1 \\ 
T_1[5,6,5,3,2] = 1 & T_1[5,6,5,4,1] = -1 & T_1[5,6,5,5,0] = 2 \\ 
\hline \hline
\caption{Tensor $T_1$ (multiplied by $24$).}
\label{Table:T1}
\end{longtable}


\begin{longtable}{MMM}
\hline \hline
T_2[0,0,0,5,6] = -1 & T_2[0,0,0,6,5] = -1 & T_2[0,0,1,4,6] = -1 \\ 
T_2[0,0,1,6,4] = 2 & T_2[0,0,2,3,6] = 1 & T_2[0,0,2,6,3] = -2 \\ 
T_2[0,0,3,2,6] = 1 & T_2[0,0,3,6,2] = -2 & T_2[0,0,4,1,6] = -1 \\ 
T_2[0,0,4,6,1] = 2 & T_2[0,0,5,0,6] = 2 & T_2[0,0,5,6,0] = -1 \\ 
T_2[0,0,6,0,5] = 2 & T_2[0,0,6,1,4] = -1 & T_2[0,0,6,2,3] = 1 \\ 
T_2[0,0,6,3,2] = 1 & T_2[0,0,6,4,1] = -1 & T_2[0,0,6,5,0] = -1 \\ 
T_2[0,1,0,4,6] = 2 & T_2[0,1,0,6,4] = -1 & T_2[0,1,4,0,6] = -1 \\ 
T_2[0,1,4,6,0] = -1 & T_2[0,1,6,0,4] = -1 & T_2[0,1,6,4,0] = 2 \\ 
T_2[0,2,0,3,6] = -2 & T_2[0,2,0,6,3] = 1 & T_2[0,2,3,0,6] = 1 \\ 
T_2[0,2,3,6,0] = 1 & T_2[0,2,6,0,3] = 1 & T_2[0,2,6,3,0] = -2 \\ 
T_2[0,3,0,2,6] = -2 & T_2[0,3,0,6,2] = 1 & T_2[0,3,2,0,6] = 1 \\ 
T_2[0,3,2,6,0] = 1 & T_2[0,3,6,0,2] = 1 & T_2[0,3,6,2,0] = -2 \\ 
T_2[0,4,0,1,6] = 2 & T_2[0,4,0,6,1] = -1 & T_2[0,4,1,0,6] = -1 \\ 
T_2[0,4,1,6,0] = -1 & T_2[0,4,6,0,1] = -1 & T_2[0,4,6,1,0] = 2 \\ 
T_2[0,5,0,0,6] = -1 & T_2[0,5,0,6,0] = 2 & T_2[0,5,6,0,0] = -1 \\ 
T_2[0,6,0,0,5] = -1 & T_2[0,6,0,1,4] = -1 & T_2[0,6,0,2,3] = 1 \\ 
T_2[0,6,0,3,2] = 1 & T_2[0,6,0,4,1] = -1 & T_2[0,6,0,5,0] = 2 \\ 
T_2[0,6,1,0,4] = 2 & T_2[0,6,1,4,0] = -1 & T_2[0,6,2,0,3] = -2 \\ 
T_2[0,6,2,3,0] = 1 & T_2[0,6,3,0,2] = -2 & T_2[0,6,3,2,0] = 1 \\ 
T_2[0,6,4,0,1] = 2 & T_2[0,6,4,1,0] = -1 & T_2[0,6,5,0,0] = -1 \\ 
T_2[1,0,1,5,6] = -2 & T_2[1,0,1,6,5] = 1 & T_2[1,0,5,1,6] = 1 \\ 
T_2[1,0,5,6,1] = 1 & T_2[1,0,6,1,5] = 1 & T_2[1,0,6,5,1] = -2 \\ 
T_2[1,1,0,5,6] = 1 & T_2[1,1,0,6,5] = -2 & T_2[1,1,1,4,6] = 1 \\ 
T_2[1,1,1,6,4] = 1 & T_2[1,1,2,3,6] = 1 & T_2[1,1,2,6,3] = -2 \\ 
T_2[1,1,3,2,6] = 1 & T_2[1,1,3,6,2] = -2 & T_2[1,1,4,1,6] = -2 \\ 
T_2[1,1,4,6,1] = 1 & T_2[1,1,5,0,6] = 1 & T_2[1,1,5,6,0] = -2 \\ 
T_2[1,1,6,0,5] = 1 & T_2[1,1,6,1,4] = -2 & T_2[1,1,6,2,3] = 1 \\ 
T_2[1,1,6,3,2] = 1 & T_2[1,1,6,4,1] = 1 & T_2[1,1,6,5,0] = 1 \\ 
T_2[1,2,1,3,6] = -2 & T_2[1,2,1,6,3] = 1 & T_2[1,2,3,1,6] = 1 \\ 
T_2[1,2,3,6,1] = 1 & T_2[1,2,6,1,3] = 1 & T_2[1,2,6,3,1] = -2 \\ 
T_2[1,3,1,2,6] = -2 & T_2[1,3,1,6,2] = 1 & T_2[1,3,2,1,6] = 1 \\ 
T_2[1,3,2,6,1] = 1 & T_2[1,3,6,1,2] = 1 & T_2[1,3,6,2,1] = -2 \\ 
T_2[1,4,1,1,6] = 1 & T_2[1,4,1,6,1] = -2 & T_2[1,4,6,1,1] = 1 \\ 
T_2[1,5,0,1,6] = 1 & T_2[1,5,0,6,1] = 1 & T_2[1,5,1,0,6] = -2 \\ 
T_2[1,5,1,6,0] = 1 & T_2[1,5,6,0,1] = -2 & T_2[1,5,6,1,0] = 1 \\ 
T_2[1,6,0,1,5] = -2 & T_2[1,6,0,5,1] = 1 & T_2[1,6,1,0,5] = 1 \\ 
T_2[1,6,1,1,4] = 1 & T_2[1,6,1,2,3] = 1 & T_2[1,6,1,3,2] = 1 \\ 
T_2[1,6,1,4,1] = -2 & T_2[1,6,1,5,0] = 1 & T_2[1,6,2,1,3] = -2 \\ 
T_2[1,6,2,3,1] = 1 & T_2[1,6,3,1,2] = -2 & T_2[1,6,3,2,1] = 1 \\ 
T_2[1,6,4,1,1] = 1 & T_2[1,6,5,0,1] = 1 & T_2[1,6,5,1,0] = -2 \\ 
T_2[2,0,2,5,6] = -2 & T_2[2,0,2,6,5] = 1 & T_2[2,0,5,2,6] = 1 \\ 
T_2[2,0,5,6,2] = 1 & T_2[2,0,6,2,5] = 1 & T_2[2,0,6,5,2] = -2 \\ 
T_2[2,1,2,4,6] = 2 & T_2[2,1,2,6,4] = -1 & T_2[2,1,4,2,6] = -1 \\ 
T_2[2,1,4,6,2] = -1 & T_2[2,1,6,2,4] = -1 & T_2[2,1,6,4,2] = 2 \\ 
T_2[2,2,0,5,6] = 1 & T_2[2,2,0,6,5] = -2 & T_2[2,2,1,4,6] = -1 \\ 
T_2[2,2,1,6,4] = 2 & T_2[2,2,2,3,6] = -1 & T_2[2,2,2,6,3] = -1 \\ 
T_2[2,2,3,2,6] = 2 & T_2[2,2,3,6,2] = -1 & T_2[2,2,4,1,6] = -1 \\ 
T_2[2,2,4,6,1] = 2 & T_2[2,2,5,0,6] = 1 & T_2[2,2,5,6,0] = -2 \\ 
T_2[2,2,6,0,5] = 1 & T_2[2,2,6,1,4] = -1 & T_2[2,2,6,2,3] = 2 \\ 
T_2[2,2,6,3,2] = -1 & T_2[2,2,6,4,1] = -1 & T_2[2,2,6,5,0] = 1 \\ 
T_2[2,3,2,2,6] = -1 & T_2[2,3,2,6,2] = 2 & T_2[2,3,6,2,2] = -1 \\ 
T_2[2,4,1,2,6] = -1 & T_2[2,4,1,6,2] = -1 & T_2[2,4,2,1,6] = 2 \\ 
T_2[2,4,2,6,1] = -1 & T_2[2,4,6,1,2] = 2 & T_2[2,4,6,2,1] = -1 \\ 
T_2[2,5,0,2,6] = 1 & T_2[2,5,0,6,2] = 1 & T_2[2,5,2,0,6] = -2 \\ 
T_2[2,5,2,6,0] = 1 & T_2[2,5,6,0,2] = -2 & T_2[2,5,6,2,0] = 1 \\ 
T_2[2,6,0,2,5] = -2 & T_2[2,6,0,5,2] = 1 & T_2[2,6,1,2,4] = 2 \\ 
T_2[2,6,1,4,2] = -1 & T_2[2,6,2,0,5] = 1 & T_2[2,6,2,1,4] = -1 \\ 
T_2[2,6,2,2,3] = -1 & T_2[2,6,2,3,2] = 2 & T_2[2,6,2,4,1] = -1 \\ 
T_2[2,6,2,5,0] = 1 & T_2[2,6,3,2,2] = -1 & T_2[2,6,4,1,2] = -1 \\ 
T_2[2,6,4,2,1] = 2 & T_2[2,6,5,0,2] = 1 & T_2[2,6,5,2,0] = -2 \\ 
T_2[3,0,3,5,6] = -2 & T_2[3,0,3,6,5] = 1 & T_2[3,0,5,3,6] = 1 \\ 
T_2[3,0,5,6,3] = 1 & T_2[3,0,6,3,5] = 1 & T_2[3,0,6,5,3] = -2 \\ 
T_2[3,1,3,4,6] = 2 & T_2[3,1,3,6,4] = -1 & T_2[3,1,4,3,6] = -1 \\ 
T_2[3,1,4,6,3] = -1 & T_2[3,1,6,3,4] = -1 & T_2[3,1,6,4,3] = 2 \\ 
T_2[3,2,3,3,6] = -1 & T_2[3,2,3,6,3] = 2 & T_2[3,2,6,3,3] = -1 \\ 
T_2[3,3,0,5,6] = 1 & T_2[3,3,0,6,5] = -2 & T_2[3,3,1,4,6] = -1 \\ 
T_2[3,3,1,6,4] = 2 & T_2[3,3,2,3,6] = 2 & T_2[3,3,2,6,3] = -1 \\ 
T_2[3,3,3,2,6] = -1 & T_2[3,3,3,6,2] = -1 & T_2[3,3,4,1,6] = -1 \\ 
T_2[3,3,4,6,1] = 2 & T_2[3,3,5,0,6] = 1 & T_2[3,3,5,6,0] = -2 \\ 
T_2[3,3,6,0,5] = 1 & T_2[3,3,6,1,4] = -1 & T_2[3,3,6,2,3] = -1 \\ 
T_2[3,3,6,3,2] = 2 & T_2[3,3,6,4,1] = -1 & T_2[3,3,6,5,0] = 1 \\ 
T_2[3,4,1,3,6] = -1 & T_2[3,4,1,6,3] = -1 & T_2[3,4,3,1,6] = 2 \\ 
T_2[3,4,3,6,1] = -1 & T_2[3,4,6,1,3] = 2 & T_2[3,4,6,3,1] = -1 \\ 
T_2[3,5,0,3,6] = 1 & T_2[3,5,0,6,3] = 1 & T_2[3,5,3,0,6] = -2 \\ 
T_2[3,5,3,6,0] = 1 & T_2[3,5,6,0,3] = -2 & T_2[3,5,6,3,0] = 1 \\ 
T_2[3,6,0,3,5] = -2 & T_2[3,6,0,5,3] = 1 & T_2[3,6,1,3,4] = 2 \\ 
T_2[3,6,1,4,3] = -1 & T_2[3,6,2,3,3] = -1 & T_2[3,6,3,0,5] = 1 \\ 
T_2[3,6,3,1,4] = -1 & T_2[3,6,3,2,3] = 2 & T_2[3,6,3,3,2] = -1 \\ 
T_2[3,6,3,4,1] = -1 & T_2[3,6,3,5,0] = 1 & T_2[3,6,4,1,3] = -1 \\ 
T_2[3,6,4,3,1] = 2 & T_2[3,6,5,0,3] = 1 & T_2[3,6,5,3,0] = -2 \\ 
T_2[4,0,4,5,6] = -2 & T_2[4,0,4,6,5] = 1 & T_2[4,0,5,4,6] = 1 \\ 
T_2[4,0,5,6,4] = 1 & T_2[4,0,6,4,5] = 1 & T_2[4,0,6,5,4] = -2 \\ 
T_2[4,1,4,4,6] = 1 & T_2[4,1,4,6,4] = -2 & T_2[4,1,6,4,4] = 1 \\ 
T_2[4,2,3,4,6] = 1 & T_2[4,2,3,6,4] = 1 & T_2[4,2,4,3,6] = -2 \\ 
T_2[4,2,4,6,3] = 1 & T_2[4,2,6,3,4] = -2 & T_2[4,2,6,4,3] = 1 \\ 
T_2[4,3,2,4,6] = 1 & T_2[4,3,2,6,4] = 1 & T_2[4,3,4,2,6] = -2 \\ 
T_2[4,3,4,6,2] = 1 & T_2[4,3,6,2,4] = -2 & T_2[4,3,6,4,2] = 1 \\ 
T_2[4,4,0,5,6] = 1 & T_2[4,4,0,6,5] = -2 & T_2[4,4,1,4,6] = -2 \\ 
T_2[4,4,1,6,4] = 1 & T_2[4,4,2,3,6] = 1 & T_2[4,4,2,6,3] = -2 \\ 
T_2[4,4,3,2,6] = 1 & T_2[4,4,3,6,2] = -2 & T_2[4,4,4,1,6] = 1 \\ 
T_2[4,4,4,6,1] = 1 & T_2[4,4,5,0,6] = 1 & T_2[4,4,5,6,0] = -2 \\ 
T_2[4,4,6,0,5] = 1 & T_2[4,4,6,1,4] = 1 & T_2[4,4,6,2,3] = 1 \\ 
T_2[4,4,6,3,2] = 1 & T_2[4,4,6,4,1] = -2 & T_2[4,4,6,5,0] = 1 \\ 
T_2[4,5,0,4,6] = 1 & T_2[4,5,0,6,4] = 1 & T_2[4,5,4,0,6] = -2 \\ 
T_2[4,5,4,6,0] = 1 & T_2[4,5,6,0,4] = -2 & T_2[4,5,6,4,0] = 1 \\ 
T_2[4,6,0,4,5] = -2 & T_2[4,6,0,5,4] = 1 & T_2[4,6,1,4,4] = 1 \\ 
T_2[4,6,2,3,4] = 1 & T_2[4,6,2,4,3] = -2 & T_2[4,6,3,2,4] = 1 \\ 
T_2[4,6,3,4,2] = -2 & T_2[4,6,4,0,5] = 1 & T_2[4,6,4,1,4] = -2 \\ 
T_2[4,6,4,2,3] = 1 & T_2[4,6,4,3,2] = 1 & T_2[4,6,4,4,1] = 1 \\ 
T_2[4,6,4,5,0] = 1 & T_2[4,6,5,0,4] = 1 & T_2[4,6,5,4,0] = -2 \\ 
T_2[5,0,5,5,6] = -1 & T_2[5,0,5,6,5] = 2 & T_2[5,0,6,5,5] = -1 \\ 
T_2[5,1,4,5,6] = -1 & T_2[5,1,4,6,5] = -1 & T_2[5,1,5,4,6] = 2 \\ 
T_2[5,1,5,6,4] = -1 & T_2[5,1,6,4,5] = 2 & T_2[5,1,6,5,4] = -1 \\ 
T_2[5,2,3,5,6] = 1 & T_2[5,2,3,6,5] = 1 & T_2[5,2,5,3,6] = -2 \\ 
T_2[5,2,5,6,3] = 1 & T_2[5,2,6,3,5] = -2 & T_2[5,2,6,5,3] = 1 \\ 
T_2[5,3,2,5,6] = 1 & T_2[5,3,2,6,5] = 1 & T_2[5,3,5,2,6] = -2 \\ 
T_2[5,3,5,6,2] = 1 & T_2[5,3,6,2,5] = -2 & T_2[5,3,6,5,2] = 1 \\ 
T_2[5,4,1,5,6] = -1 & T_2[5,4,1,6,5] = -1 & T_2[5,4,5,1,6] = 2 \\ 
T_2[5,4,5,6,1] = -1 & T_2[5,4,6,1,5] = 2 & T_2[5,4,6,5,1] = -1 \\ 
T_2[5,5,0,5,6] = 2 & T_2[5,5,0,6,5] = -1 & T_2[5,5,1,4,6] = -1 \\ 
T_2[5,5,1,6,4] = 2 & T_2[5,5,2,3,6] = 1 & T_2[5,5,2,6,3] = -2 \\ 
T_2[5,5,3,2,6] = 1 & T_2[5,5,3,6,2] = -2 & T_2[5,5,4,1,6] = -1 \\ 
T_2[5,5,4,6,1] = 2 & T_2[5,5,5,0,6] = -1 & T_2[5,5,5,6,0] = -1 \\ 
T_2[5,5,6,0,5] = -1 & T_2[5,5,6,1,4] = -1 & T_2[5,5,6,2,3] = 1 \\ 
T_2[5,5,6,3,2] = 1 & T_2[5,5,6,4,1] = -1 & T_2[5,5,6,5,0] = 2 \\ 
T_2[5,6,0,5,5] = -1 & T_2[5,6,1,4,5] = -1 & T_2[5,6,1,5,4] = 2 \\ 
T_2[5,6,2,3,5] = 1 & T_2[5,6,2,5,3] = -2 & T_2[5,6,3,2,5] = 1 \\ 
T_2[5,6,3,5,2] = -2 & T_2[5,6,4,1,5] = -1 & T_2[5,6,4,5,1] = 2 \\ 
T_2[5,6,5,0,5] = 2 & T_2[5,6,5,1,4] = -1 & T_2[5,6,5,2,3] = 1 \\ 
T_2[5,6,5,3,2] = 1 & T_2[5,6,5,4,1] = -1 & T_2[5,6,5,5,0] = -1 \\ 
\hline \hline
\caption{Tensor $T_2$ (multiplied by $12\sqrt{5}$).}
\label{Table:T2}
\end{longtable}


\begin{longtable}{MMM}
\hline \hline
T_3[0,0,0,5,6] = -1 & T_3[0,0,0,6,5] = 1 & T_3[0,0,1,4,6] = 1 \\ 
T_3[0,0,2,3,6] = -1 & T_3[0,0,3,2,6] = -1 & T_3[0,0,4,1,6] = 1 \\ 
T_3[0,0,5,6,0] = -1 & T_3[0,0,6,1,4] = -1 & T_3[0,0,6,2,3] = 1 \\ 
T_3[0,0,6,3,2] = 1 & T_3[0,0,6,4,1] = -1 & T_3[0,0,6,5,0] = 1 \\ 
T_3[0,1,0,6,4] = -1 & T_3[0,1,4,0,6] = -1 & T_3[0,1,4,6,0] = 1 \\ 
T_3[0,1,6,0,4] = 1 & T_3[0,2,0,6,3] = 1 & T_3[0,2,3,0,6] = 1 \\ 
T_3[0,2,3,6,0] = -1 & T_3[0,2,6,0,3] = -1 & T_3[0,3,0,6,2] = 1 \\ 
T_3[0,3,2,0,6] = 1 & T_3[0,3,2,6,0] = -1 & T_3[0,3,6,0,2] = -1 \\ 
T_3[0,4,0,6,1] = -1 & T_3[0,4,1,0,6] = -1 & T_3[0,4,1,6,0] = 1 \\ 
T_3[0,4,6,0,1] = 1 & T_3[0,5,0,0,6] = 1 & T_3[0,5,6,0,0] = -1 \\ 
T_3[0,6,0,0,5] = -1 & T_3[0,6,0,1,4] = 1 & T_3[0,6,0,2,3] = -1 \\ 
T_3[0,6,0,3,2] = -1 & T_3[0,6,0,4,1] = 1 & T_3[0,6,1,4,0] = -1 \\ 
T_3[0,6,2,3,0] = 1 & T_3[0,6,3,2,0] = 1 & T_3[0,6,4,1,0] = -1 \\ 
T_3[0,6,5,0,0] = 1 & T_3[1,0,1,6,5] = 1 & T_3[1,0,5,1,6] = 1 \\ 
T_3[1,0,5,6,1] = -1 & T_3[1,0,6,1,5] = -1 & T_3[1,1,0,5,6] = -1 \\ 
T_3[1,1,1,4,6] = 1 & T_3[1,1,1,6,4] = -1 & T_3[1,1,2,3,6] = -1 \\ 
T_3[1,1,3,2,6] = -1 & T_3[1,1,4,6,1] = 1 & T_3[1,1,5,0,6] = -1 \\ 
T_3[1,1,6,0,5] = 1 & T_3[1,1,6,2,3] = 1 & T_3[1,1,6,3,2] = 1 \\ 
T_3[1,1,6,4,1] = -1 & T_3[1,1,6,5,0] = 1 & T_3[1,2,1,6,3] = 1 \\ 
T_3[1,2,3,1,6] = 1 & T_3[1,2,3,6,1] = -1 & T_3[1,2,6,1,3] = -1 \\ 
T_3[1,3,1,6,2] = 1 & T_3[1,3,2,1,6] = 1 & T_3[1,3,2,6,1] = -1 \\ 
T_3[1,3,6,1,2] = -1 & T_3[1,4,1,1,6] = -1 & T_3[1,4,6,1,1] = 1 \\ 
T_3[1,5,0,1,6] = 1 & T_3[1,5,0,6,1] = -1 & T_3[1,5,1,6,0] = 1 \\ 
T_3[1,5,6,1,0] = -1 & T_3[1,6,0,5,1] = 1 & T_3[1,6,1,0,5] = -1 \\ 
T_3[1,6,1,1,4] = 1 & T_3[1,6,1,2,3] = -1 & T_3[1,6,1,3,2] = -1 \\ 
T_3[1,6,1,5,0] = -1 & T_3[1,6,2,3,1] = 1 & T_3[1,6,3,2,1] = 1 \\ 
T_3[1,6,4,1,1] = -1 & T_3[1,6,5,0,1] = 1 & T_3[2,0,2,6,5] = 1 \\ 
T_3[2,0,5,2,6] = 1 & T_3[2,0,5,6,2] = -1 & T_3[2,0,6,2,5] = -1 \\ 
T_3[2,1,2,6,4] = -1 & T_3[2,1,4,2,6] = -1 & T_3[2,1,4,6,2] = 1 \\ 
T_3[2,1,6,2,4] = 1 & T_3[2,2,0,5,6] = -1 & T_3[2,2,1,4,6] = 1 \\ 
T_3[2,2,2,3,6] = -1 & T_3[2,2,2,6,3] = 1 & T_3[2,2,3,6,2] = -1 \\ 
T_3[2,2,4,1,6] = 1 & T_3[2,2,5,0,6] = -1 & T_3[2,2,6,0,5] = 1 \\ 
T_3[2,2,6,1,4] = -1 & T_3[2,2,6,3,2] = 1 & T_3[2,2,6,4,1] = -1 \\ 
T_3[2,2,6,5,0] = 1 & T_3[2,3,2,2,6] = 1 & T_3[2,3,6,2,2] = -1 \\ 
T_3[2,4,1,2,6] = -1 & T_3[2,4,1,6,2] = 1 & T_3[2,4,2,6,1] = -1 \\ 
T_3[2,4,6,2,1] = 1 & T_3[2,5,0,2,6] = 1 & T_3[2,5,0,6,2] = -1 \\ 
T_3[2,5,2,6,0] = 1 & T_3[2,5,6,2,0] = -1 & T_3[2,6,0,5,2] = 1 \\ 
T_3[2,6,1,4,2] = -1 & T_3[2,6,2,0,5] = -1 & T_3[2,6,2,1,4] = 1 \\ 
T_3[2,6,2,2,3] = -1 & T_3[2,6,2,4,1] = 1 & T_3[2,6,2,5,0] = -1 \\ 
T_3[2,6,3,2,2] = 1 & T_3[2,6,4,1,2] = -1 & T_3[2,6,5,0,2] = 1 \\ 
T_3[3,0,3,6,5] = 1 & T_3[3,0,5,3,6] = 1 & T_3[3,0,5,6,3] = -1 \\ 
T_3[3,0,6,3,5] = -1 & T_3[3,1,3,6,4] = -1 & T_3[3,1,4,3,6] = -1 \\ 
T_3[3,1,4,6,3] = 1 & T_3[3,1,6,3,4] = 1 & T_3[3,2,3,3,6] = 1 \\ 
T_3[3,2,6,3,3] = -1 & T_3[3,3,0,5,6] = -1 & T_3[3,3,1,4,6] = 1 \\ 
T_3[3,3,2,6,3] = -1 & T_3[3,3,3,2,6] = -1 & T_3[3,3,3,6,2] = 1 \\ 
T_3[3,3,4,1,6] = 1 & T_3[3,3,5,0,6] = -1 & T_3[3,3,6,0,5] = 1 \\ 
T_3[3,3,6,1,4] = -1 & T_3[3,3,6,2,3] = 1 & T_3[3,3,6,4,1] = -1 \\ 
T_3[3,3,6,5,0] = 1 & T_3[3,4,1,3,6] = -1 & T_3[3,4,1,6,3] = 1 \\ 
T_3[3,4,3,6,1] = -1 & T_3[3,4,6,3,1] = 1 & T_3[3,5,0,3,6] = 1 \\ 
T_3[3,5,0,6,3] = -1 & T_3[3,5,3,6,0] = 1 & T_3[3,5,6,3,0] = -1 \\ 
T_3[3,6,0,5,3] = 1 & T_3[3,6,1,4,3] = -1 & T_3[3,6,2,3,3] = 1 \\ 
T_3[3,6,3,0,5] = -1 & T_3[3,6,3,1,4] = 1 & T_3[3,6,3,3,2] = -1 \\ 
T_3[3,6,3,4,1] = 1 & T_3[3,6,3,5,0] = -1 & T_3[3,6,4,1,3] = -1 \\ 
T_3[3,6,5,0,3] = 1 & T_3[4,0,4,6,5] = 1 & T_3[4,0,5,4,6] = 1 \\ 
T_3[4,0,5,6,4] = -1 & T_3[4,0,6,4,5] = -1 & T_3[4,1,4,4,6] = -1 \\ 
T_3[4,1,6,4,4] = 1 & T_3[4,2,3,4,6] = 1 & T_3[4,2,3,6,4] = -1 \\ 
T_3[4,2,4,6,3] = 1 & T_3[4,2,6,4,3] = -1 & T_3[4,3,2,4,6] = 1 \\ 
T_3[4,3,2,6,4] = -1 & T_3[4,3,4,6,2] = 1 & T_3[4,3,6,4,2] = -1 \\ 
T_3[4,4,0,5,6] = -1 & T_3[4,4,1,6,4] = 1 & T_3[4,4,2,3,6] = -1 \\ 
T_3[4,4,3,2,6] = -1 & T_3[4,4,4,1,6] = 1 & T_3[4,4,4,6,1] = -1 \\ 
T_3[4,4,5,0,6] = -1 & T_3[4,4,6,0,5] = 1 & T_3[4,4,6,1,4] = -1 \\ 
T_3[4,4,6,2,3] = 1 & T_3[4,4,6,3,2] = 1 & T_3[4,4,6,5,0] = 1 \\ 
T_3[4,5,0,4,6] = 1 & T_3[4,5,0,6,4] = -1 & T_3[4,5,4,6,0] = 1 \\ 
T_3[4,5,6,4,0] = -1 & T_3[4,6,0,5,4] = 1 & T_3[4,6,1,4,4] = -1 \\ 
T_3[4,6,2,3,4] = 1 & T_3[4,6,3,2,4] = 1 & T_3[4,6,4,0,5] = -1 \\ 
T_3[4,6,4,2,3] = -1 & T_3[4,6,4,3,2] = -1 & T_3[4,6,4,4,1] = 1 \\ 
T_3[4,6,4,5,0] = -1 & T_3[4,6,5,0,4] = 1 & T_3[5,0,5,5,6] = 1 \\ 
T_3[5,0,6,5,5] = -1 & T_3[5,1,4,5,6] = -1 & T_3[5,1,4,6,5] = 1 \\ 
T_3[5,1,5,6,4] = -1 & T_3[5,1,6,5,4] = 1 & T_3[5,2,3,5,6] = 1 \\ 
T_3[5,2,3,6,5] = -1 & T_3[5,2,5,6,3] = 1 & T_3[5,2,6,5,3] = -1 \\ 
T_3[5,3,2,5,6] = 1 & T_3[5,3,2,6,5] = -1 & T_3[5,3,5,6,2] = 1 \\ 
T_3[5,3,6,5,2] = -1 & T_3[5,4,1,5,6] = -1 & T_3[5,4,1,6,5] = 1 \\ 
T_3[5,4,5,6,1] = -1 & T_3[5,4,6,5,1] = 1 & T_3[5,5,0,6,5] = -1 \\ 
T_3[5,5,1,4,6] = 1 & T_3[5,5,2,3,6] = -1 & T_3[5,5,3,2,6] = -1 \\ 
T_3[5,5,4,1,6] = 1 & T_3[5,5,5,0,6] = -1 & T_3[5,5,5,6,0] = 1 \\ 
T_3[5,5,6,0,5] = 1 & T_3[5,5,6,1,4] = -1 & T_3[5,5,6,2,3] = 1 \\ 
T_3[5,5,6,3,2] = 1 & T_3[5,5,6,4,1] = -1 & T_3[5,6,0,5,5] = 1 \\ 
T_3[5,6,1,4,5] = -1 & T_3[5,6,2,3,5] = 1 & T_3[5,6,3,2,5] = 1 \\ 
T_3[5,6,4,1,5] = -1 & T_3[5,6,5,1,4] = 1 & T_3[5,6,5,2,3] = -1 \\ 
T_3[5,6,5,3,2] = -1 & T_3[5,6,5,4,1] = 1 & T_3[5,6,5,5,0] = -1 \\ 
\hline \hline
\caption{Tensor $T_3$ (multiplied by $4\sqrt{15}$).}
\label{Table:T3}
\end{longtable}




\bibliography{bib_SU4_Z2}

